\documentclass[authoryear,preprint]{elsarticle}

\usepackage{amssymb}
\usepackage{amsmath}
\usepackage{bm}
\usepackage{amsthm}
\usepackage{amsfonts}
\usepackage{hyperref}
\hypersetup{
colorlinks = true,
linkcolor = cyan,
filecolor = blue,
urlcolor = red,
citecolor = green,
}
\usepackage[shadow,loadshadowlibrary,textsize=small,textwidth=2.5cm]{todonotes}
\usepackage{booktabs}
\usepackage{graphicx}
\usepackage{subcaption}
\usepackage{algorithm,algorithmic}
\usepackage{geometry}
\usepackage{tablefootnote}
\usepackage{booktabs}
\usepackage{dashundergaps}
\usepackage{etoolbox}
\usepackage{nomencl}
\newtheorem{remark}{Remark}
\makenomenclature
\biboptions{sort&compress}

\usepackage{lineno}
\usepackage{mathtools}
\newcommand{\lsp}{\bm{s}^\mathrm{p}}
\newcommand{\lspd}{\bm{s}^\mathrm{p}_\mathrm{dev}}
\newcommand{\epsp}{\bm{\varepsilon}^\mathrm{p}}
\newcommand{\eps}{\bm{\varepsilon}}

\newcommand{\mrm}[1]{{\mathrm{#1}}}

\begin{document}

\begin{frontmatter}



\title{
Revisiting the hydromechanical formulation of a micromechanics-based phase-field model for poro-elastoplastic media
}



\author[label1,label2]{Hanzhang Li} 
\ead{hanzhang_li@tongji.edu.cn}
\author[label3,label4]{Tao You\corref{cor1}}
\ead{tao.you@unileoben.ac.at}
\author[label3,label4]{Keita Yoshioka}
\ead{keita.yoshioka@unileoben.ac.at}
\author[label1,label2]{Yuhao Liu}
\ead{2111030@tongji.edu.cn}
\author[label1,label2]{Yi Rui}
\ead{ruiyi@tongji.edu.cn}
\author[label1,label2]{Fengshou Zhang\corref{cor1}}
\ead{fengshou.zhang@tongji.edu.cn}

\affiliation[label1]{organization={State Key Laboratory of Disaster Reduction in Civil Engineering, Tongji University},
            addressline={1239 Siping Road}, 
            city={Shanghai},
            postcode={200092}, 
            country={China}}
\affiliation[label2]{organization={Department of Geotechnical Engineering, College of Civil Engineering, Tongji University},
            addressline={1239 Siping Road}, 
            city={Shanghai},
            postcode={200092}, 
            country={China}}
\affiliation[label3]{organization={Department Geoenergy, Montanuniversit\"at Leoben},
            addressline={Parkstra\ss e 27}, 
            city={Leoben},
            postcode={8700}, 
            country={Austria}}
            
\affiliation[label4]{organization={Department of Environmental Informatics, Helmholtz Centre for Environmental Research - UFZ},
            addressline={Permoserstra\ss e 15}, 
            city={Leipzig},
            postcode={04318}, 
            country={Germany}}
\cortext[cor1]{Corresponding author}
\begin{abstract}
Even for tension-dominated fracture propagation, porous materials may deform plastically adjacent to the propagating fracture. 
As is common for porous materials, existing phase-field models typically employ a non-associative flow rule for plasticity, and a Helmholtz free energy based on strain and fluid pressure.
This work revisits the hydromechanically coupled formulation of the phase-field model for fracture in poro-elastoplastic media by analyzing the strength surface and fracture driving force. Our analyses show that these common choices of flow rule and free energy will lead to a discontinuous strength surface across the tension-compression transition. A non-associative flow rule introduces a jump at the strength surface, while treating fluid pressure—rather than fluid content—as the independent variable in the Helmholtz free energy omits a coupling term from the phase-field driving force, also breaking continuity. 
Incorporating an associative Drucker–Prager flow rule and this omitted coupling term ensures a continuous strength surface and the accurate fracture driving force. The proposed model exhibits improved accuracy in hydromechanical responses when compared against the analytical solution of the Kristianovich–Geertsma–de Klerk hydraulic fracturing benchmark.
Numerical simulations of hydraulic fracturing and biaxial compression in poro-elastoplastic media show that the model can reproduce both shear-dominated fractures induced by mechanical disturbance and tension-dominated fractures driven by fluid injection in saturated porous media. 


\end{abstract}




\end{frontmatter}




\section{Introduction}
\label{Introduction}


While the fracturing of porous media is governed solely by the solid skeleton in the dry state, in fully or partially saturated states, it is driven by both the skeleton deformation and the fluid flow in the pores and defects. 
Fluid-saturated fracturing in porous media has received significant focus, not only due to industrial applications—such as enhanced production in oil and gas reservoirs~\citep{economides2000reservoir} and the creation of enhanced geothermal systems~\citep{horne2025enhanced}—but also because of environmentally assisted fracturing (such as the carbonation of cementitious materials~\citep{korec2024predicting}) and geological events such as iceberg crevassing~\citep{mobasher2016modeling} and dyke intrusions~\citep{mori2022three}.

In modeling hydraulic fracturing, early foundational studies assumed elasticity to derive analytical solutions~\citep{kristianovitch1955formation, 10.2118/2458-PA, garagash2006plane, detournay2016mechanics, barenblatt1956formation}. 
Similarly, most classical numerical models have relied on linear elastic fracture mechanics~\citep{adachi2007computer, clifton1979computation, clifton1989three, sousa1993numerical, shah1997hydraulic}. 
To overcome the limitations of discrete crack tracking, phase-field (or gradient damage) formulations have been proposed~\citep{bourdin2000numerical, bourdin2008variational, miehe2010phase} and successfully applied to various fracture problems, such as ductile~\citep{ALESSI2015351} and fatigue fracturing~\citep{Carrara2020}.
Phase-field models have also been extended to coupled hydromechanical problems~\citep{bourdin2012variational, miehe2015minimization, yoshioka2016variational, wheeler2020ipacs}, where fluid mass evolution is typically governed by a mass balance equation incorporating Darcy's law. 
Meanwhile, the total energy functional is modified to incorporate pore pressure effects. 
As summarized in \citep{you_poroelastic_2023}, the exact form of this modified total energy varies across the literature. 
The predominant practice is to treat pore pressure $p$ directly as the independent variable in the Helmholtz free energy functional \citep{heider2020phase, zhou2019phase, chukwudozie2019variational, yoshioka2016variational}.\footnote{Throughout this paper, we distinguish \textit{independent variables} from \textit{primary variables}. Independent variables are the arguments of a thermodynamic potential with respect to which partial differentiation is performed---e.g., $\bm{\varepsilon}$ and $\xi$ for the Helmholtz free energy, or $\bm{\sigma}$ and $p$ for the Gibbs free energy. Primary variables, in contrast, refer to the nodal unknowns $(\bm{u}, p, d)$ solved in the finite element formulation.}
Alternatively, several studies employ the fluid content variation $\xi$ as the independent variable \citep{ULLOA2022115084, mikelic2015phase}\footnote{Note that \citet{mikelic2015phase} initially formulates the systematic free energy functional in terms of fluid content $\xi$ before substituting it with pressure $p$, and a similar procedure is presented in \citet{you_poroelastic_2023}.}. 
From a thermodynamic standpoint, the free energy of the solid matrix–fluid system can be formulated via either Helmholtz energy or Gibbs energy. 
A Helmholtz free energy naturally takes strain $\bm{\varepsilon}$, fluid content $\xi$, and internal damage variable $\omega$ as its canonical independent variables. 
Conversely, a Gibbs free energy takes stress $\bm{\sigma}$, fluid pressure $p$, and $\omega$ as independent variables.
In practice, however, most existing literature adopts a ``mixed'' formulation, applying the Helmholtz free energy in which fluid content $\xi$ is replaced by pressure $p$. 
While this mixed formulation offers computational convenience, as pressure appears directly in boundary conditions and is immediately interpretable, the Legendre transform between the Helmholtz and Gibbs free energies no longer holds. 
This inconsistency in the Legendre transformation may lead to oversights when deriving generalized driving forces.
Though such omissions rarely cause numerical divergence, they will compromise the accuracy of the resulting mechanical responses.

Another gap in the current hydromechanical phase-field models is that existing models have predominantly focused on crack nucleation and propagation in purely poroelastic media~\citep{Heider2017, santillan2017phase, chukwudozie2019variational}.
Under realistic conditions involving high fluid pressures and elevated confining stresses, plastic deformation can become significant. 
For mechanically induced failure in poro-elastoplastic media, micromechanics-based damage models incorporating strength criteria offer the flexibility needed to capture fracture nucleation and propagation~\citep{dormieux2006microporomechanics, dormieux2007micromechanics, XIE2012919, ZHU2023103789}. For instance, \citet{ZHU2023103789} presented experimental evidence and a micromechanical damage model incorporating a Coulomb friction criterion to explain shear fracturing in saturated quasi-brittle rocks. 
Despite these advances, the coupled effect of plastic deformation on hydraulic fracture growth in quasi-brittle materials remains insufficiently explored.

To frame the analytical investigation in this paper, we briefly restate the widely accepted concepts of fracture nucleation and the strength surface. 
In the phase-field modeling framework, fracture nucleation is defined as damage localization within narrow bands ($\Delta d \neq 0$)~\citep{kumar2020phase, vicentini2024energy, lopez2025classical}. 
Experimentally, for a given material, the peak (or critical) stresses under a spatially uniform, monotonically increasing, but otherwise arbitrary loading process represent the onset of fracture nucleation; the locus of these critical stresses in stress space defines the strength surface of the material. 
Note that certain strength surfaces derived from phase-field models~\citep{vicentini2024energy, khayaz2025comparisonphasefieldmodels} can only capture fracture nucleation under pure tension or specific loading states (e.g., uniaxial tensile and compressive strengths), as discussed in~\citet{lopez2025classical, li2025cohesive}.



In this paper, we aim to revisit the hydromechanical coupled formulation and provide a more rigorous derivation. 
First, we take poroelasticity, poroplasticity~\citep{Coussy}, and microporomechanics~\citep{you_poroelastic_2023, ULLOA2022115084} as our starting point. 
The Helmholtz free energy is defined using strain $\bm{\varepsilon}$, fluid content $\xi$, and internal variables $d$ (phase-field variable) as independent variables. 
We use a representative volume element (RVE) to determine the state of microcracks and, accordingly, derive the generalized stresses under tension and compression. Then, we employ the modified cohesive-type degradation function proposed in our previous work~\citep{li2025cohesive} for the phase-field evolution. Finally, we consider a Drucker–Prager-type strength criterion incorporating an associative flow rule to model plastic flow.

This paper is structured as follows. 
Section~\ref{Micromechanical-based phase field framework} presents a micromechanics-based phase-field model for hydromechanical fracture. 
Section~\ref{sec:ss} derives the strength surfaces from the strength criterion and phase-field driving force of the proposed model, comparing them against those derived from existing models. 
Section~\ref{sec: numerical implementation} outlines the numerical implementation procedure, including the finite element discretization of the governing equations and the solution algorithm for the three-field ($d$-$p$-$\bm{u}$) problem.
Section~\ref{Numerical experiment} demonstrates numerical examples, including hydraulic fracturing in poroelastic and poro-elastoplastic materials as well as biaxial compression in saturated poro-elastoplastic media. 
Finally, Section~\ref{Conclusion} summarizes the main conclusions.

\section{Model formulation}
\label{Micromechanical-based phase field framework}

\begin{figure}[h!]
\centering
\includegraphics[scale=0.4]{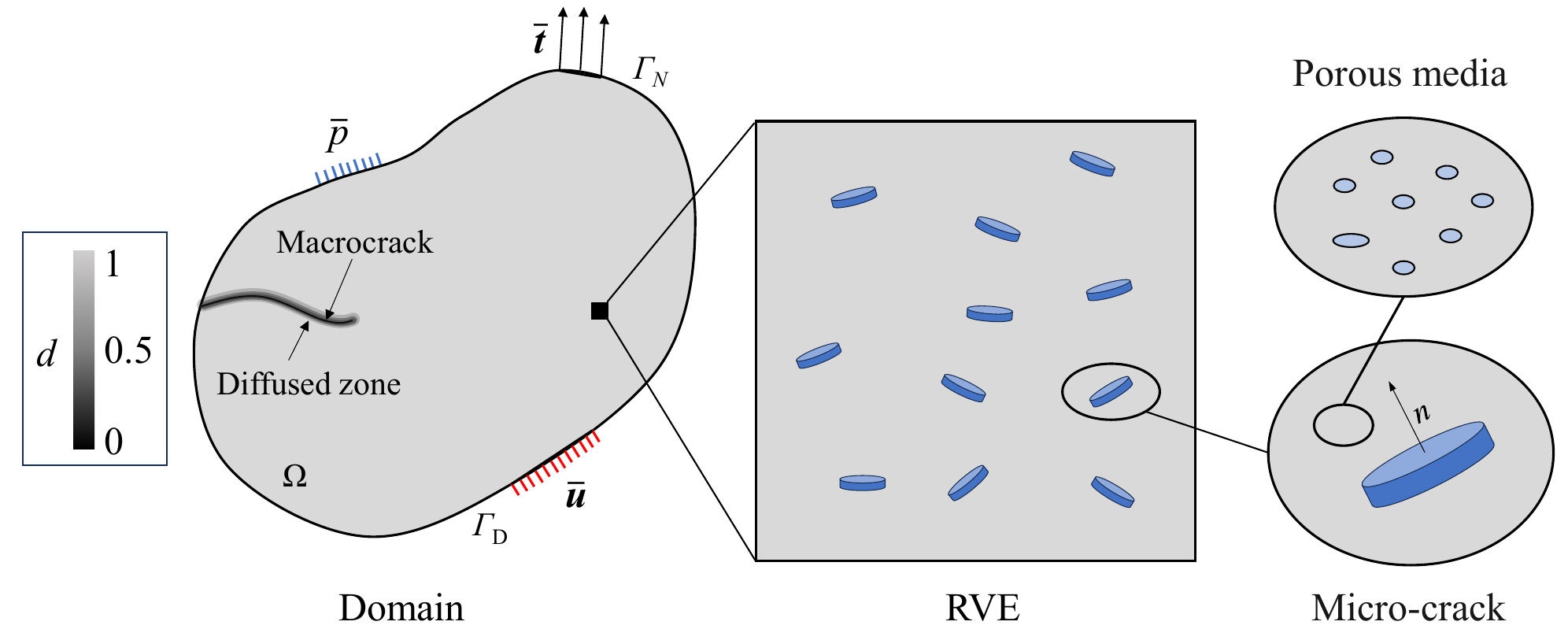}
\caption{A study object at two scales, where a phase-field scalar $d \in [0,1]$ is introduced for the regularization of the macrocrack (left), while the penny-shaped microcrack (right) is randomly distributed in the saturated RVE (middle). }
\label{fig: research body}
\end{figure}

In this section, we derive the cohesive-frictional phase-field model in saturated porous media to include pore fluid interactions with the solid skeleton based on~\citep{li2025cohesive}.  
Consider a fully saturated domain $\Omega \subset \mathbb{R}^{n_\mathrm{dim}}$ ($n_\mathrm{dim} = 2,3$) subjected to traction $\bar{\mathbf{t}}$, displacement $\bar{\mathbf{u}}$, and pressure $\bar{p}$ on the boundary. 
A representative volume element (RVE) is extracted to depict the randomly distributed penny-shaped microcracks incorporating numerous fluid-filled pores at the microscale, as illustrated in Fig.~\ref{fig: research body}. Depending on the stress state and fluid pressure, the microcracks can open or close.
At the macroscale, the fracture is described using the phase-field variable, $d \in [0,1]$, where $d=0$ represents fully damaged state and $d=1$ the intact state. 

\subsection{Free energy of the system}
Building on previous analyses of microcrack behavior within the RVE~\citep{zhu2011micromechanics, XIE2012919, you_novel_2021}, we focus on two distinct states: crack opening under tensile conditions, and crack closure accompanied by frictional sliding under compressive/shear conditions. 
The strain tensor ($\boldsymbol{\varepsilon}$) can be decomposed as
\begin{equation}
    \label{total strain}
    \bm{\varepsilon} = \bm{\varepsilon}^\mathrm{e} + \bm{\varepsilon}^\mathrm{p}
    ,
\end{equation}
where $\bm{\varepsilon}^\mathrm{e}$ is the elastic strain and $\bm{\varepsilon}^\mathrm{p}$ is the inelastic strain induced by either microcrack opening or frictional sliding of closed microcracks. 

In porous media, the fluid content $\xi$ participates in the interaction between the solid skeleton and the pore fluid.
For fully saturated quasi-brittle materials, it takes the form:
\begin{equation}
\label{eq: fluid content}
    \xi(\bm{\varepsilon},d) = \phi - \phi_0,
\end{equation}
where $\phi$ and $\phi_0$ denote the current and reference Lagrangian porosities. 
Following~\citet{Coussy}, we introduce two referenced hydromechanical properties - the reference Biot coefficient $\alpha_0$ and Biot modulus $M_0$:
\begin{equation}
    \begin{aligned}
        \alpha_0 &= 1 - \frac{K}{K_\mathrm{s}}, \\
        \frac{1}{M_0} &= \frac{\alpha_0 - \phi_0}{K_\mathrm{s}} + \phi_0 c_f
        ,
    \end{aligned}
\end{equation}
where $K$ is the bulk modulus of the material, $K_\mathrm{s}$ is the bulk modulus of the skeleton, and $c_f$ is the compressibility coefficient of the fluid. 
Their evolutions are related to the phase-field variable~\citep{XIE2012919, ULLOA2022115084, you_poroelastic_2023} as
\begin{equation}
    \begin{aligned}
        \alpha &= \alpha_0 + (1 - \alpha_0)[1 - g(d)], \\
        \frac{1}{M} &= \frac{\alpha - \phi_0}{K_\mathrm{s}} + \phi_0 c_f \\
        &= \frac{1}{M_0} + [1 - g(d)]\frac{(1 - \alpha_0)^2}{K},
    \end{aligned}
    \label{eq:biot}
\end{equation}
with $g(d)$ being the degradation function.  

Following~\cite{francfort1998revisiting}, we define the total energy of the system as
\begin{equation}
    \Psi_\mathrm{total} = \Psi - \Psi_\mathrm{ext} = \Psi - \int_\Omega \bm{b} \cdot \bm{u} \mathrm{d}V - \int_{\Gamma_N} \bar{\bm{t}} \cdot \bm{u} \,\mathrm{d}S
    ,
\end{equation}
where $\Psi$ is the Helmholtz free energy.
$\Psi$ is composed of the bulk energy $\Psi_b$ and the crack surface energy $\Psi_s$ as
\begin{equation}
    \Psi = \Psi_\mathrm{b} + \Psi_\mathrm{s} =\int_{\Omega}\psi(\bm{\varepsilon},\xi, d)\mathrm{d}V + \int_{\Omega}G_c \gamma(d,\nabla d)\mathrm{d}V
    ,
    \label{eq: free energy}
\end{equation}
where $\psi(\bm{\varepsilon}, \xi, d)$ denotes the bulk energy density in porous media, and $\gamma(d, \nabla d)$ is the crack surface density expressed by the phase-field variable $d$. 

To represent these two distinct mechanisms, we define the bulk energy $\psi(\bm{\varepsilon},\xi, d)$ as follows: 
\begin{enumerate}
    \item[$\bullet$] $\texttt{Open microcracks}$. In this case, the inelastic strain $\bm{\varepsilon}^\mathrm{p}$ is induced by microcrack opening. 
    The free energy density of the RVE is given based on the poromechanics-based model~\citep{Coussy, mikelic2015phase, you_poroelastic_2023} as
    \begin{equation}
        \label{homo open strain energy}
        \psi_{\mathrm{open}}(\bm{\varepsilon}, \xi, d) = \frac{1}{2}\bm{\varepsilon}:\mathbb{C}_{\mathrm{dam}}(d):\bm{\varepsilon} + \frac{M}{2}\left( \alpha \mathrm{tr}[\bm{\varepsilon}] - \xi\right)^2
        ,
    \end{equation}
   where $\mathbb{C}_{\mathrm{dam}} (d)$ is the degraded elastic tensor defined as 
   $$\mathbb{C}_{\mathrm{dam}} (d) = g(d)\mathbb{C}.$$
   $\mathbb{C}$ is the elastic tensor of the intact material defined as $\mathbb{C} = 3K\mathbb{J} + 2\mu\mathbb{K}$ where $K$ and $\mu$ denote bulk modulus and shear modulus, $\mathbb{J}$ and $\mathbb{K}$ denote fourth-order identity tensor and deviatoric tensor.
    
    \item[$\bullet$] $\texttt{Closed microcracks}$. The frictional sliding of closed microcracks contributes to the inelastic strain $\bm{\varepsilon}^\mathrm{p}$. The free energy density can be written as
    \begin{equation}
        \label{homo closed strain energy}
        \psi_{\mathrm{close}}(\bm{\varepsilon}, \bm{\varepsilon}^\mathrm{p}, \xi , d) = \frac{1}{2}(\bm{\varepsilon}-\bm{\varepsilon}^\mathrm{p}):\mathbb{C}:(\bm{\varepsilon}-\bm{\varepsilon}^\mathrm{p}) + \frac{1}{2}\bm{\varepsilon}^\mathrm{p}:\mathbb{H}(d):\bm{\varepsilon}^\mathrm{p} + \frac{M_0}{2}\left( \alpha_0 \mathrm{tr}[\bm{\varepsilon} - \bm{\varepsilon}^\mathrm{p}] + \mathrm{tr}[\bm{\varepsilon}^\mathrm{p}] - \xi\right)^2
        ,
    \end{equation}
   where the first term of the right side denotes the elastic strain energy, the second term is the inelastic strain energy related to microcracks, and the third term is the contribution of fluid-to-solid coupling to the free energy. The fourth-order tensor $\mathbb{H}(d)$ is the kinematic hardening modulus.
\end{enumerate}

The pressure $p$ is derived by taking derivative of $\psi$ related to the fluid content $\xi$~\citep{Coussy, ULLOA2022115084}, which reads
\begin{equation}
    \label{eq:pressure}
    p(\bm{\varepsilon}, \bm{\varepsilon}^\mathrm{p}, \xi , d) = \frac{\partial \psi}{\partial \xi} =
    \begin{cases}
        -M\left( \alpha \mathrm{tr}[\bm{\varepsilon}] - \xi\right) &\quad \texttt{Open microcracks} \\
        -M_0\left( \alpha_0 \mathrm{tr}[\bm{\varepsilon} - \bm{\varepsilon}^\mathrm{p}] + \mathrm{tr}[\bm{\varepsilon}^\mathrm{p}] - \xi\right) &\quad \texttt{Closed microcracks}
    \end{cases}
    .
\end{equation}

Employing the Coleman-Noll procedure, the stress-strain relations for these two cases can be derived:
\begin{equation}
    \label{stress-strain relation}
    \begin{aligned}
        \bm{\sigma}_{\mathrm{open}} = \frac{\partial\psi_{\mathrm{open}}}{\partial\bm{\varepsilon}} =& \mathbb{C}_{\mathrm{dam}}(d):\bm{\varepsilon} + \alpha M \left( \alpha \mathrm{tr}[\bm{\varepsilon}] - \xi \right) \mathbf{I} \\
        =& \mathbb{C}_{\mathrm{dam}}(d):\bm{\varepsilon} - \alpha p \mathbf{I}, \\
        \bm{\sigma}_{\mathrm{close}} = \frac{\partial\psi_{\mathrm{close}}}{\partial\bm{\varepsilon}} =& \mathbb{C}:(\bm{\varepsilon}-\bm{\varepsilon}^\mathrm{p}) + \alpha_0 M_0 \left( \alpha_0 \mathrm{tr}[\bm{\varepsilon} - \bm{\varepsilon}^\mathrm{p}] + \mathrm{tr}[\bm{\varepsilon}^\mathrm{p}] - \xi\right) \mathbf{I} \\
        =& \mathbb{C}:(\bm{\varepsilon}-\bm{\varepsilon}^\mathrm{p}) - \alpha_0 p \mathbf{I}
    .
    \end{aligned}
\end{equation}

The generalized stresses conjugate to the inelastic strain $\boldsymbol{\varepsilon}^\mathrm{p}$ are given as
\begin{equation}
    \label{generalized stress inelastic}
    \begin{aligned}
        \bm{s}_{\mathrm{open}}^{\mathrm{p}} = -\frac{\partial\psi_{\mathrm{open}}}{\partial\bm{\varepsilon}^\mathrm{p}} =& 0, \\
        \bm{s}_{\mathrm{close}}^{\mathrm{p}} = -\frac{\partial\psi_{\mathrm{close}}}{\partial\bm{\varepsilon}^\mathrm{p}} =& \mathbb{C}:(\bm{\varepsilon}-\bm{\varepsilon}^\mathrm{p}) - \mathbb{H}(d):\bm{\varepsilon}^\mathrm{p} - (1 - \alpha_0)M_0\left( \alpha_0 \mathrm{tr}[\bm{\varepsilon} - \bm{\varepsilon}^\mathrm{p}] + \mathrm{tr}[\bm{\varepsilon}^\mathrm{p}] - \xi\right)\mathbf{I} \\
        =& \mathbb{C}:(\bm{\varepsilon}-\bm{\varepsilon}^\mathrm{p}) - \mathbb{H}(d):\bm{\varepsilon}^\mathrm{p} + (1 - \alpha_0)p \mathbf{I}
    .
    \end{aligned}
\end{equation}

The generalized stresses conjugate to the phase-field variable take the form\footnote{
\citet{ULLOA2022115084} propose a similar driving force at microcrack opening as 
$$
\bm{s}_{\mathrm{open}}^{\mathrm{d}}
        = -\frac{1}{2}\bm{\varepsilon}:[3Kg^\prime_K(d)\mathbb{J} + 2\mu g^\prime_\mu(d)\mathbb{K}]:\bm{\varepsilon} - \frac{(1 - \alpha_0)^2}{2K}p^2g^\prime_K(d) - (1 - \alpha_0)pg^\prime_K(d)\mathrm{tr}[\bm{\varepsilon}]
        .
$$
where the difference is that~\citet{ULLOA2022115084} employs $g_K(d)$ and $g_\mu(d)$ individually for the volumetric and deviatoric parts of the elastic tensors, while we use a unified degradation function.
}
\begin{equation}
    \begin{aligned}
        s_{\mathrm{open}}^{\mathrm{d}} = -\frac{\partial\psi_{\mathrm{open}}}{\partial d} = & -\frac{1}{2}\bm{\varepsilon}:\mathbb{C}^\prime_{\mathrm{dam}}(d):\bm{\varepsilon} - \frac{\partial}{\partial d}\left[\frac{M}{2}\left( \alpha \mathrm{tr}[\bm{\varepsilon}] - \xi\right)^2 \right]     \\
        & = -\frac{1}{2}g^\prime(d)\bm{\varepsilon}:\mathbb{C}:\bm{\varepsilon} - \frac{(1 - \alpha_0)^2}{2K}p^2g^\prime(d) - (1 - \alpha_0)pg^\prime(d)\mathrm{tr}[\bm{\varepsilon}], \\
        s_{\mathrm{close}}^{\mathrm{d}} = -\frac{\partial\psi_{\mathrm{close}}}{\partial d} = & -\frac{1}{2}\bm{\varepsilon}^\mathrm{p}:\mathbb{H}^\prime(d):\bm{\varepsilon}^\mathrm{p}
    .
    \end{aligned}
    \label{generalized stress density}
\end{equation}
To arrive at Eq.~\eqref{generalized stress density}-1, we applied the partial derivatives of $\alpha$ and $M$ with respect to $d$:
\begin{equation}
    \label{eq:pd of M}
    \begin{aligned}
        \frac{\partial \alpha}{\partial d} =& -(1 - \alpha_0)g^\prime(d) \\
        \frac{\partial M}{\partial d} =& - \frac{K}{(1 - \alpha_0)}\frac{\alpha ^\prime}{(\alpha - \phi_0)^2} = M^2 \frac{(1 - \alpha_0)^2}{K}g^\prime(d)
    \end{aligned}
    ,
\end{equation}
and Eq.~\eqref{eq:pressure}-1 for $\texttt{Open microcracks}$.
\begin{remark}
The poroelastic strain energy $\psi$ takes the strain $\bm{\varepsilon}$, the fluid content $\xi$, and the phase-field $d$ as independent variables, and the inelastic strain $\epsp$ as an internal variable.
To this end, the derived variables -- stress $\bm{\sigma}$ and pressure $p$ -- should be viewed as functions with respect to the independent variables and the internal variables.
\end{remark}

The continuity of stress and energy at the opening and closure transition of microcracks requires:
\begin{equation}
    \label{continuous}
    \bm{\sigma}_{\mathrm{close}} = \bm{\sigma}_{\mathrm{open}}, \quad
    \bm{s}_{\mathrm{close}}^{\mathrm{p}} = \bm{s}_{\mathrm{open}}^{\mathrm{p}}, \quad
    s_{\mathrm{close}}^{\mathrm{d}} = s_{\mathrm{open}}^{\mathrm{d}},
    \quad
    \psi_\mathrm{open} = \psi_\mathrm{close}
    .
\end{equation}
Combining Eq.~\eqref{continuous}--1 and Eq.~\eqref{continuous}--2, we have
\begin{equation}
\label{block tensor}
    \mathbb{H}(d) = g_p(d)\mathbb{C}
    ,
\end{equation}
where 
\begin{equation}
    g_p(d) = \frac{g(d)}{1 - g(d)}
    \label{eq: gp}
\end{equation}
denotes the degradation function of the hardening modulus (See \ref{Consistent hardening modulus} for the derivation).
Combining Eqs.~\eqref{continuous}--2, \eqref{generalized stress inelastic} and \eqref{block tensor}, we obtain the relation between plastic strain and total strain for open microcracks, i.e.,
\begin{equation}
    \bm{\varepsilon}^\mathrm{p} = [1 - g(d)]\bm{\varepsilon} + \mathbb{C}^{-1}:(\alpha - \alpha_0)p\mathbf{I}.
    \label{eq: ep open}
\end{equation}

Substituting Eq.~\eqref{eq: ep open} into Eq.~\eqref{homo closed strain energy} results in Eq.~\eqref{homo open strain energy}, confirming that Eq.~\eqref{homo closed strain energy} provides a unified definition of the free energy for both closed and open microcracks. 
Similarly to our previous work in~\cite{li2025cohesive}, we designate  $$\boldsymbol{s}^\mathrm{p}=\mathbb{C}:(\bm{\varepsilon - \epsp}) - \mathbb{H}(d):\bm{\varepsilon}^\mathrm{p} + (1 - \alpha_0)p\mathbf{I}$$ as a unified expression for the generalized stress with respect to the inelastic strain $\boldsymbol{\varepsilon}^\mathrm{p}$. Then, under the isotropic assumption, the microcrack opening-closure indicator can be defined as
\begin{align}
   \label{trace of sp}
   \begin{aligned}
    \begin{cases}
    \mathrm{tr}[\bm{s}^{\mathrm{p}}]=0  & \mathrm{for} \quad \texttt{Open microcracks}  \\
        \mathrm{tr}[\bm{s}^{\mathrm{p}}]<0  & \mathrm{for} \quad \texttt{Closed microcracks}
    \end{cases}
   \end{aligned}
   .
\end{align}



\subsection{Evolution of phase-field}
The degradation function $g(d)$ and crack surface density function $\gamma(d,\nabla d)$ from the cohesive phase-field model originally proposed by~\citet{lorentz2011convergence, WU201772} are given as:
\begin{align}
    \label{eq: g(d)}
    g(d) &= \frac{d^{m}}{d^{m}+a_1(1 - d)[1 + a_2 (1 - d) + a_3 (1 - d)^2]} \\
    \gamma(d,\nabla d) &= \frac{1}{\pi} \left(\frac{1 - d^{2}}{\ell} + \ell|\nabla d|^{2} \right)
    ,
\end{align}
where $\ell$ is the phase-field length scale parameter. 
The coefficients $m$, $a_2$, and $a_3$ in Eq.~\eqref{eq: g(d)} can be selected from a specific softening curve. Here, we employ a linear softening law so that $m = 2$, $a_2 = - \frac{1}{2}$, and $a_3 = 0$. 
The cohesive coefficient $a_1$ takes the form 
\begin{equation}
\label{eq:a1}
    a_1 = \frac{4l_\mathrm{ch}}{\pi \ell}.
\end{equation}

The internal length $l_\mathrm{ch}$ is principal-stress dependent as proposed in~\citet{li2025cohesive}:
\begin{equation}
\label{stress dependent characteristic length}
    \begin{aligned}
        l_\mathrm{ch} =
        \begin{cases}
            l_\mathrm{t} &\mathrm{for} \quad \sigma_{\mathrm{sup}} < \sigma_\mathrm{min} \\
            [1 - g_{s}(\sigma_\mathrm{min})]l_\mathrm{c} + g_{s}(\sigma_\mathrm{min})l_\mathrm{t} &\mathrm{for} \quad \sigma_{\mathrm{inf}} < \sigma_\mathrm{min} \leq \sigma_{\mathrm{sup}} \\
            l_\mathrm{c} \quad &\mathrm{for} \quad \sigma_\mathrm{min} \leq \sigma_{\mathrm{inf}}
        \end{cases}
    \end{aligned}
\end{equation}
with $\sigma_\mathrm{min} = \mathrm{min}\{ \sigma_1, \, \sigma_2, \, \sigma_3\}$ the minimum principal stress. 
$l_\mathrm{t}$ is the characteristic length for tensile fracture given by Irwin's definition as $$l_\mathrm{t} = \frac{G_c E}{f^2_\mathrm{t}},$$ where $E$ is the Young's modulus, and $f_\mathrm{t}$ is the uniaxial tensile strength. 
$l_\mathrm{c}$ is the characteristic length for compressive failure and can be determined by fitting the experimental strength surface. 
The transition function $g_{s}(\sigma_\mathrm{min})$ takes the following form:
\begin{equation}
\label{eq: gs}
    g_{s}(\sigma_\mathrm{min}) = \frac{\eta^2}{\eta^2 + \kappa(1 - \eta^2)} \quad \mathrm{with} \quad \eta = \frac{\sigma_\mathrm{min} - \sigma_{\mathrm{inf}}}{\sigma_{\mathrm{sup}} - \sigma_{\mathrm{inf}}}
    ,
\end{equation}
where $\sigma_{\mathrm{inf}}$ and $\sigma_{\mathrm{sup}}$ are the material parameters to be calibrated from experiment and denote the lower and upper bounds of the transition point. $\kappa$ is a constant parameter that regulates the transition profile.

In line with the general variational framework presented in~\citet{ ULLOA2022115084}, the evolution of the phase-field is expressed as
\begin{equation}
\label{evolution criterion of pf}
    e^{d}(d) = -\frac{\partial\psi(\bm{\varepsilon}, \bm{\varepsilon}^\mathrm{p}, \xi, d)}{\partial d} - G_c \delta_d \gamma(d, \nabla d)  \le 0 , \quad d \in \mathcal{D}(t_{n - 1})
\end{equation}
where the variational symbol is defined as $\delta_d \square := \partial_d \square - \nabla\cdot\partial_{\nabla d}\square$, and $t_{n-1}$ denotes the last pseudo time. The admissible set of $d$ is defined as
\begin{equation}
    \mathcal{D}(t_{n-1}) = \{ d \in H^1 (\Omega): 0 \leq d(x, t_{n}) \leq d(x, t_{n - 1}) \leq 1 \quad \forall x \, \, \mathrm{s.t.} \, \, d(x, t_{n - 1}) \leq d_\mathrm{irr} \}
\end{equation}
where $d_\mathrm{irr} \in [0,1]$ is the irreversible threshold introduced in~\cite{bourdin2007numerical, burke2013adaptive}. 
For $d_\mathrm{irr} = 1$, we recover the strict irreversible condition.

Based on the generalized stress conjugate for different states of microcracks Eq.~\eqref{generalized stress density}, we can expand Eq.~\eqref{evolution criterion of pf} as
\begin{align}
\label{eq: evolution of phase field}
  \begin{aligned}
      e^{d}(d) &=
    \begin{dcases}
        -\frac{1}{2}\frac{\partial g(d)}{\partial d} \left( \bm{\varepsilon}:\mathbb{C}:\bm{\varepsilon} + \frac{(1 - \alpha_0)^2}{K}p^2 + 2(1 - \alpha_0)p\mathrm{tr}[\bm{\varepsilon}] \right) + \dfrac{2G_{c}}{\pi} \left( \frac{d}{\ell} + \ell \Delta d \right) &\mathrm{for}   \quad \mathrm{tr}[\bm{s}^{\mathrm{p}}]=0 \\
        -\frac{1}{2}\dfrac{\partial g_{p}(d)}{\partial d}\bm{\varepsilon}^{\mathrm{p}}:\mathbb{C}:\bm{\varepsilon}^{\mathrm{p}} + \dfrac{2G_{c}}{\pi} \left( \frac{d}{\ell} + \ell \Delta d \right)   &\mathrm{for} \quad \mathrm{tr}[\bm{s}^{\mathrm{p}}]<0
    \end{dcases}
  \end{aligned}
\end{align}
where the first derivatives of the degradation functions $g(d)$ and $g_{p}(d)$ are:
\begin{equation}
    \begin{aligned}
        \frac{\partial g(d)}{\partial d} &= \frac{a_1 d}{\left(d^{2}+\frac{1}{2}a_{1}(1 - d^{2})\right)^2}, \\
        \frac{\partial g_p(d)}{\partial d} &= \frac{\partial}{\partial d}\left(\frac{g(d)}{1 - g(d)}\right) = \frac{g'(d)}{\left(1 - g(d)\right)^{2}}. \\
    \end{aligned}
\end{equation}

\begin{remark}
While an identical poroelastic strain energy functional was proposed in previous works~\citep{mikelic2015phase, yi2020consistent,you_poroelastic_2023}, those formulations adopt the strain $\boldsymbol{\varepsilon}$ and fluid pressure $p$ as independent variables. 
This particular choice of state variables yields the phase-field evolution for open microcracks ($\mathrm{tr}[\bm{s}^{\mathrm{p}}]=0$) as
\begin{equation}
\label{eq: uncoupled phase-field evolution}
     e^d_\mathrm{mixed}  = -\frac{1}{2}\frac{\partial g(d)}{\partial d} \left[ \bm{\varepsilon}:\mathbb{C}:\bm{\varepsilon} + \frac{(1 - \alpha_0)^2}{K}p^2  \right] + \dfrac{2G_{c}}{\pi} \left( \frac{d}{\ell} + \ell \Delta d \right)
     .
\end{equation}
where we call this kind of model the "mixed formulation". Comparing Eq.~\eqref{eq: uncoupled phase-field evolution} against Eq.~\eqref{eq: evolution of phase field}-1 reveals that the coupling term, $ 2(1 - \alpha_0)p\mathrm{tr}[\bm{\varepsilon}]$, is omitted from the first bracket.
This omission leads to a discontinuous strength surface, which will be elaborated in Sec.~\ref{sec:ss}.  
\end{remark}

\subsection{Evolution of plasticity}


To evaluate the frictional sliding, we employ a Drucker-Prager-type friction criterion incorporating an associative flow rule (AFR) in the local stress space as

\begin{equation}
    \label{yield func}
    \begin{aligned}
       \text{Friction criterion} \quad& f^{\mathrm{p}}(\bm{s}^{\mathrm{p}}) := \| \bm{s}^{\mathrm{p}}_{\mathrm{dev}} \| + \frac{A}{3} \mathrm{tr} [\bm{s}^{\mathrm{p}}]\le 0 \\
       \text{Plastic potential} \quad& h^\mathrm{p}(\bm{s}^{\mathrm{p}}) := \| \bm{s}^{\mathrm{p}}_{\mathrm{dev}} \| + \frac{A}{3} \mathrm{tr} [\bm{s}^{\mathrm{p}}]
    \end{aligned}
\end{equation}
where $\bm{s}^{\mathrm{p}}_{\mathrm{dev}} = \mathbb{K}:\bm{s}^\mathrm{p}$ is the deviatoric part of $\bm{s}^\mathrm{p}$, and $A$ denotes the frictional coefficient of closed microcracks. 

At time step $n$, given the strain increment $\Delta \bm{\varepsilon}$, $\boldsymbol{\varepsilon}^\mathrm{p}$ is obtained from
\begin{align}
\label{incremental}
    \begin{dcases}
        \bm{\varepsilon}^\mathrm{p} = \bm{\varepsilon}^\mathrm{p}_{n-1} + \Delta \bm{\varepsilon}^\mathrm{p} = \bm{\varepsilon}^\mathrm{p}_{n-1} + \Delta \lambda \mathbf{D}, 
        \\
        \bm{s}^\mathrm{p} = \bm{s}^\mathrm{p \, trial} - \Delta\lambda(\mathbb{C} + \mathbb{H}(d)):\mathbf{D}, \\
        \mathbf{D} =  \frac{\partial h^\mathrm{p}(\bm{s}^\mathrm{p})}{\partial \bm{s}^\mathrm{p}} = \mathbf{V} + \frac{A}{3}\mathbf{I} \quad \text{with} \quad \mathbf{V} := \frac{\bm{s}^\mathrm{p}_\mathrm{dev}}{\| \bm{s}^\mathrm{p}_\mathrm{dev} \|}\\
        f^\mathrm{p}(\bm{s}^{\mathrm{p}}) \leq 0, \quad \Delta \lambda \geq 0, \quad \Delta \lambda f^\mathrm{p}(\bm{s}^{\mathrm{p}}) = 0
    \end{dcases}
\end{align}
where $\mathbf{D}$ denotes the flow direction tensor, $\mathbf{V}$ is the deviatoric part of $\mathbf{D}$, and $\Delta \lambda$ is the incremental Lagrangian multiplier. 
$\bm{s}^\mathrm{p \, trial}$ is the trial state of the local stress $\bm{s}^\mathrm{p}$, which reads
\begin{equation}
\label{eq:trsp}
    \begin{aligned}
        \bm{s}^\mathrm{p \, trial} &:= \mathbb{C}:(\bm{\varepsilon} - \bm{\varepsilon}^\mathrm{p}_{n-1}) - \mathbb{H}(d)\bm{\varepsilon}^\mathrm{p}_{n-1} + (1 - \alpha_0)p\mathbf{I} \\
        &= \bm{\sigma}^\mathrm{trial} - \mathbb{H}(d)\bm{\varepsilon}^\mathrm{p}_{n-1} + p\mathbf{I}
    \end{aligned}
    .
\end{equation}

The equivalent plastic strain ${\varepsilon}^\mathrm{p,eq}$ is introduced as the measure of frictional sliding in compressive-shear fracture, and its rate $\dot{{\varepsilon}}^\mathrm{p,eq} $ can be obtained from the following equation
\begin{equation}
    \dot{{\varepsilon}}^\mathrm{p,eq} =
    \begin{cases}
        0  & \quad \mathrm{for} \quad \mathrm{tr}[\bm{s}^\mathrm{p}] = 0 ,\\
        \sqrt{2/3} \| \dot{\bm{\varepsilon}}^\mathrm{p} \|  & \quad \mathrm{for} \quad \mathrm{tr}[\bm{s}^\mathrm{p}] < 0.
    \end{cases}
\end{equation}

\subsection{Fluid flow model}
The mass balance is given with the Darcy flow in porous media as
\begin{equation}
\label{eq: fluid ge}
    \begin{dcases}
        \dfrac{\partial \xi}{\partial t} + \nabla\cdot \left(- \dfrac{\bm{K}}{\nu}\nabla p \right) = Q \quad &\text{in} \quad \Omega  \\
        p = \bar{p} \quad &\text{on} \quad \partial\Omega_p \\
        \bm{\upsilon}\cdot\mathbf{n} = \bar{\bm{q}} \quad &\text{on} \quad \partial\Omega_q
    \end{dcases}
\end{equation}
where $Q$ denotes the source/sink term, $\bm{K}$ denotes the permeability tensor, and $\nu$ denotes the fluid viscosity. $\bar{p}$ and $\bar{\bm{q}}$ denote the prescribe pressure and normal flux, respectively. 
For the enhanced permeability in the fractured porous media, we employ the following formulation~\citep{you2026dual}:
\begin{equation}
    \bm{K} = K_\mathrm{m}\mathbf{I} + (1 - d)^{k}\frac{w^3}{12h}(\mathbf{I} - \mathbf{n}_\Gamma\otimes\mathbf{n}_\Gamma)
\end{equation}
where $K_\mathrm{m}$ denotes the isotropic permeability of the porous media, $k \geq 1$ denotes an additional material parameter, and $w$ denotes the fracture aperture calculated by~\citep{miehe2015minimization}
\begin{equation}
    w = \mathrm{tr}[\bm{\varepsilon}]h
    ,
\end{equation}
where $h$ is the characteristic length and is taken as the element size in this work.

The normal vector of the crack is obtained from the eigenvector associated with the maximum principal strain as
\begin{equation}
    \mathbf{n}_\Gamma = \mathbf{e}_1 \quad \text{with} \quad \bm{\varepsilon} = \sum_{l = 1}^3 \varepsilon_l \mathbf{e}_l
    ,
\end{equation}
where $\varepsilon_l$ are the principal strains with $\varepsilon_1 > \varepsilon_2 > \varepsilon_3$, and $\mathbf{e}_l$ are the corresponding eigenvectors.
For the verification of fracture aperture computation by this approach, we refer to~\cite{you_poroelastic_2023}.

\section{Strength surfaces under coupled hydromechanical conditions}
\label{sec:ss}
This section investigates strength surfaces under different free energy formulations and plastic flow rules. We first derive the strength surface for the proposed model using a consistent Helmholtz free energy formulation. To validate this result, we compare it with the surface derived from the Gibbs free energy, which offers a clear definition in terms of stress $\boldsymbol{\sigma}$ and fluid pressure $p$. Following this theoretical comparison, we calibrate the model parameters by verifying the strength surface against the experimental data of ~\citep{ZHU2023103789}. We then examine the continuity of strength surfaces for the mixed Helmholtz free energy formulation. Finally, we assess the influence of the degradation function and plastic flow rule on the strength surfaces.

\subsection{Strength surface derived from the proposed model}
\label{sec: strength derivation}

The phase-field evolution, Eq.~\eqref{evolution criterion of pf}, represents the first variation of the total energy with respect to damage.
Considering homogeneous damage ($\nabla d = 0$), and the strength surface can be derived from the critical point ($e^d = 0$):

\begin{equation}
\label{eq:critical_point}
    \begin{dcases}
         -\frac{1}{2}\frac{\partial g(d)}{\partial d} \left[ \bm{\varepsilon}:\mathbb{C}:\bm{\varepsilon} + \frac{(1 - \alpha_0)^2}{K}p^2 + 2(1 - \alpha_0)p\mathrm{tr}[\bm{\varepsilon}] \right] + \dfrac{2G_{c}}{\pi} \dfrac{d}{\ell} = 0&\mathrm{for}   \quad \mathrm{tr}[\bm{s}^{\mathrm{p}}]=0 \\
        -\frac{1}{2}\dfrac{\partial g_{p}(d)}{\partial d}\bm{\varepsilon}^{\mathrm{p}}:\mathbb{C}:\bm{\varepsilon}^{\mathrm{p}} + \dfrac{2G_{c}}{\pi} \dfrac{d}{\ell}=0   &\mathrm{for} \quad \mathrm{tr}[\bm{s}^{\mathrm{p}}]<0
    \end{dcases}
    .
\end{equation}

\subsection*{Open microcracks}

Eq.~\eqref{eq:critical_point}-1 presents the strength surface for the open microcrack case.
From the total stress-strain relation $ \bm{\sigma} = g(d) \mathbb{C}:\bm{\varepsilon} - \alpha p \mathbf{I} $ and using $\mathrm{tr}[\bm{\varepsilon}] = \frac{1}{g(d)}\frac{1}{K}\left( \frac{\mathrm{tr}[\bm{\sigma}]}{3} + \alpha p \right)$, we have
\begin{equation}
\label{eq:analytical-1}
    \begin{aligned}
        0 =& \frac{g^\prime(d)}{2 g(d)^2} \left[ \frac{1}{2\mu}\bm{\sigma}_\mathrm{dev}:\bm{\sigma}_\mathrm{dev} + \frac{1}{K}\left(\sigma_\mathrm{sph} + \alpha p + (1 - \alpha_0) g(d) p \right)^2 \right] - \dfrac{2G_{c}}{\pi} \dfrac{d}{\ell}\\
        =& \frac{1}{2\mu}\bm{\sigma}_\mathrm{dev}:\bm{\sigma}_\mathrm{dev} + \frac{1}{K}\left(\sigma_\mathrm{sph} + p \right)^2 - \dfrac{G_{c} d^4}{l_\mathrm{ch}} := \mathcal{F}_\mathrm{open}(\bm{\sigma} , d)
        ,
    \end{aligned}
\end{equation}
where $\sigma_\mathrm{sph} = \frac{\mathrm{tr}[\bm{\sigma}]}{3}$ and $\bm{\sigma}_\mathrm{dev} = \bm{\sigma} - \sigma_\mathrm{sph}\mathbf{I}$ are the mean and deviatoric stresses.

Eq.~\eqref{eq:analytical-1} shows that the strength surface $\mathcal{F}_\mathrm{open}(\bm{\sigma},d)$ is $\ell$-independent and monotonic with respect to $d$ under the coupled hydromechanical condition. Given the initial phase-field value $d_0 = 1.0$, the strength surface is expressed as
\begin{equation}
\label{eq: ss_open_coupled}
    \mathcal{F}_\mathrm{open}(\bm{\sigma},d_0) = \frac{\| \bm{\sigma}_\mathrm{dev} \|^2}{2\mu} + \frac{(\sigma_\mathrm{sph} + p)^2}{K} - \frac{G_c}{l_\mathrm{ch}}
    .
\end{equation}

\subsection*{Closed microcracks}

Under a monotonic loading, the evolution of inelastic strain follows $\dot{\bm{\varepsilon}^\mathrm{p}} = \dot{\lambda} \mathbf{D}$, then we have
\begin{equation}
\label{eq: inelastic_strain}
    \bm{\varepsilon}^\mathrm{p} = \Lambda \mathbf{D}
    ,
\end{equation}
where $\Lambda = \int \dot{\lambda} \mathrm{d}t$. 
We start by transforming the friction criterion Eq.~\eqref{yield func} from the local stress space to the principal stress space. 
Recalling Eq.~\eqref{generalized stress inelastic} and Eq.~\eqref{incremental}, $\bm{s}^\mathrm{p}_\mathrm{dev}$ and $\mathrm{tr}[\bm{s}^\mathrm{p}]$ can be expressed as
\begin{equation}
\label{eq:decomposition of sp}
    \begin{aligned}
        \bm{s}^\mathrm{p}_\mathrm{dev} &= \mathbb{K}:\bm{s}^\mathrm{p} = \bm{\sigma}_\mathrm{dev} - 2\mu g_p\Lambda \frac{\bm{s}^\mathrm{p}_\mathrm{dev}}{\| \bm{s}^\mathrm{p}_\mathrm{dev} \|}, \\
        \mathrm{tr}[\bm{s}^\mathrm{p}] &= \mathbf{I}:\bm{s}^\mathrm{p} = \mathrm{tr}[\bm{\sigma}] - 3AKg_p\Lambda + 3p
        .
    \end{aligned}
\end{equation}
Thus, we have
\begin{equation}
\label{eq:norm of sp_dev}
        \bm{\sigma}_\mathrm{dev}:\bm{\sigma}_\mathrm{dev} 
        = \left(1 + 2\mu g_p\Lambda \frac{1}{\| \bm{s}^\mathrm{p}_\mathrm{dev} \|} \right)^2 \bm{s}^\mathrm{p}_\mathrm{dev}:\bm{s}^\mathrm{p}_\mathrm{dev} ,
\end{equation}
and therefore
\begin{equation}
    \| \bm{s}^\mathrm{p}_\mathrm{dev} \| = \| \bm{\sigma}_\mathrm{dev} \| - 2\mu g_p\Lambda .
\end{equation}

Substituting Eqs.~\eqref{eq:decomposition of sp}-2 and \eqref{eq:norm of sp_dev}-2 into Eq.~\eqref{yield func}, the strength surface becomes
\begin{equation}
\label{eq:yield_close}
    \| \bm{\sigma}_\mathrm{dev} \| + A(\sigma_\mathrm{sph} + p) - g_p \Lambda \chi = 0,
\end{equation}
where $\chi =\mathbf{D}:\mathbb{C}:\mathbf{D} = A^2 K + 2\mu$. 

Furthermore, using Eq.~\eqref{eq: inelastic_strain}, Eq.\eqref{eq:critical_point}-2 can be written as
\begin{equation}
\label{eq:Lambda}
    -\frac{1}{2}g^\prime_p(d) \Lambda^2 \chi 
        + \frac{2 G_c d}{\pi \ell}  = 0.
\end{equation}

Substituting Eq.~\eqref{eq:Lambda} into Eq.~\eqref{eq:yield_close} to eliminate $\Lambda$, we obtain
\begin{equation}
\label{eq:fclose}
    \begin{aligned}
        0 =& \| \bm{\sigma}_\mathrm{dev} \| + A(\sigma_\mathrm{sph} + p) - g_p \chi \sqrt{\frac{4G_c d}{\pi \ell g^\prime_p \chi}} \\
        =&  \| \bm{\sigma}_\mathrm{dev} \| + A(\sigma_\mathrm{sph} + p) - \sqrt{\frac{G_c d^4 \chi}{l_\mathrm{ch}}} := \mathcal{F}_\mathrm{close}(\bm{\sigma} , d)
        .
    \end{aligned}
\end{equation}

Also for closed microcracks, the strength surface Eq.~\eqref{eq:fclose} is independent of the phase-field length scale $\ell$, and monotonic with respect to $d$. 
With an initial value of phase-field $d_0 = 1.0$, the strength surface for intact materials is expressed as
\begin{equation}
\label{eq:ss_close}
    \mathcal{F}_\mathrm{close}(\bm{\sigma}, d_0) = \| \bm{\sigma}_\mathrm{dev} \| + A(\sigma_\mathrm{sph} + p) - \sqrt{\frac{G_c \chi}{l_\mathrm{ch}}}
    .
\end{equation}

Equation~\eqref{eq: ss_open_coupled} gives the strength surface for open microcracks. Together with Eq.~\eqref{eq:ss_close}, it defines the complete strength surface for both open and closed microcracks under pore pressure. Notably, pore pressure influences only the spherical part of the stress tensor in the strength surface, following the basic concept of the effective stress principle.

\subsection{Strength surface derived from the Gibbs free energy}
\label{app:gibbs}

To theoretically verify the above strength surface, we derive it from the Gibbs free energy. Following the Legendre-Fenchel transformation $\psi ^\star = \bm{\sigma}:\bm{\varepsilon} + p\xi - \psi$, the bulk energy density in the Gibbs free energy density is given as~\citep{ZHU2023103789}
\begin{equation}
    \begin{aligned}
        \psi_\mathrm{open}^{\star}(\bm{\sigma}, p, d) = 
            \frac{1}{2} (\bm{\sigma} + \alpha p \mathbf{I}): \mathbb{S}_\mathrm{dam}(d):(\bm{\sigma} + \alpha p \mathbf{I}) + \frac{1}{2} \frac{p^2}{M}
            ,
    \end{aligned}
\end{equation}

\begin{equation}
    \begin{aligned}
        \psi_\mathrm{close}^{\star}(\bm{\sigma}, \bm{\varepsilon}^\mathrm{p}, p, d) =& 
            \frac{1}{2}\bm{\sigma}:\mathbb{S}:\bm{\sigma} + \bm{\sigma}:\bm{\varepsilon}^\mathrm{p} - \frac{1}{2}\bm{\varepsilon}^\mathrm{p}:\mathbb{H}(d):\bm{\varepsilon}^\mathrm{p} + p \mathbf{I}:\bm{\varepsilon}^\mathrm{p} \\
            &+ \alpha_0 p \mathbf{I}:\mathbb{S}:\bm{\sigma} +\frac{1}{2}p^2 \left( \alpha_0^2 \mathbf{I}:\mathbb{S}:\mathbf{I} + \frac{1}{M_0} \right) 
            ,
    \end{aligned}
\end{equation}
where $\mathbb{S} = \mathbb{C}^{-1}$ and $\mathbb{S}_\mathrm{dam}(d) = \mathbb{C}_\mathrm{dam}^{-1}(d) = \frac{1}{g(d)}\mathbb{C}^{-1} = \frac{1}{g(d)}\mathbb{S}$ are defined as the degraded compliance tensor and intact compliance tensor.
Employing the Coleman-Noll procedure, the stress-strain relations for these two cases read
\begin{equation}
    \label{eq:gibbs_stress_strain}
    \bm{\varepsilon}_{\mathrm{open}} = \frac{\partial\psi^{\star}_{\mathrm{open}}}{\partial\bm{\sigma}} = \mathbb{S}_{\mathrm{dam}}(d):(\bm{\sigma} + \alpha p \mathbf{I}), \quad \bm{\varepsilon}_{\mathrm{close}} = \frac{\partial\psi^{\star}_{\mathrm{close}}}{\partial\bm{\sigma}} = \mathbb{S}:\bm{\sigma} + \bm{\varepsilon}^\mathrm{p} + \mathbb{S}:\alpha_0 p \mathbf{I}   .
\end{equation}

The generalized stresses conjugate to the inelastic strain $\boldsymbol{\varepsilon}_p$ are given as
\begin{equation}
    \label{eq:gibbs_local_stress}
    \bm{s}_{\mathrm{open}}^{\mathrm{p},\star} =\frac{\partial\psi^*_{\mathrm{open}}}{\partial\bm{\varepsilon}^\mathrm{p}} = 0, \quad \bm{s}_{\mathrm{close}}^{\mathrm{p}, \star} = \frac{\partial\psi^{\star}_{\mathrm{close}}}{\partial\bm{\varepsilon}^\mathrm{p}} = \bm{\sigma} - \mathbb{H}(d):\bm{\varepsilon}^\mathrm{p} + p\mathbf{I}
    .
\end{equation}

The generalized stresses conjugate to the phase-field variable take the form
\begin{equation}
    \begin{aligned}
        \bm{s}_{\mathrm{open}}^{\mathrm{d}, \star} =\frac{\partial\psi^{\star}_{\mathrm{open}}}{\partial d} =& \left[-\frac{g'(d)}{2g(d)^2}(\bm{\sigma} + \alpha p \mathbf{I}) - g'(d)(1 - \alpha_0)p\mathbf{I}  \right]:\mathbb{S}:(\bm{\sigma} + \alpha p \mathbf{I}) - g'(d)\frac{(1 - \alpha_0)^2 p^2}{2K} \\
        =& -\frac{g'(d)}{2g(d)^2}\left[ \frac{1}{2\mu}\bm{\sigma}_\mathrm{dev}:\bm{\sigma}_\mathrm{dev} + \frac{1}{K} \left( \frac{\mathrm{tr}\bm{\sigma}}{3} + p \right)^2 \right]
    , \\
    \bm{s}_{\mathrm{close}}^{\mathrm{d}, \star} =\frac{\partial\psi^{\star}_{\mathrm{close}}}{\partial d} =& -\frac{1}{2}\bm{\varepsilon}^\mathrm{p}:\mathbb{H}'(d):\bm{\varepsilon}^\mathrm{p}
    .
    \end{aligned}
    \label{eq:gibbs_driving_force}
\end{equation}

Following the same procedure presented in Section~\ref{sec: strength derivation}, the strength surface can be derived as
\begin{equation}
    \begin{aligned}
        \mathcal{F}^{\star}_\mathrm{open}(\bm{\sigma},d_0) =& \frac{\| \bm{\sigma}_\mathrm{dev} \|^2}{2\mu} + \frac{(\sigma_\mathrm{sph} + p)^2}{K} - \frac{G_c}{l_\mathrm{ch}} = 0 \\
        \mathcal{F}^{\star}_\mathrm{close}(\bm{\sigma}, d_0) =& \| \bm{\sigma}_\mathrm{dev} \| + A(\sigma_\mathrm{sph} + p) - \sqrt{\frac{G_c \chi}{l_\mathrm{ch}}} = 0
    \end{aligned}
\end{equation}
which confirms a consistent expression with the strength surface in Sec.~\ref {sec: strength derivation}.

\subsection{Comparison against experimental data}
\label{sec:comparison against experimental data}

We then verify the compressive-shear strength surface given by Eq.~\eqref{eq:ss_close} against the conventional triaxial compression experiments under coupled hydromechanical conditions, reported by~\citet{ZHU2023103789}. 
In their experiments, cylinder specimens of the gray sandstone were prepared with a diameter of~50 mm and a height of 100~mm. 
A hydrostatic confining stress was first applied to the specimen surfaces ($\sigma_1 = \sigma_2 = \sigma_3 = p_c$, where $p_c$ is the prescribed confining stress), followed by water injection into the specimens to achieve the initial pore pressure lower than the confining pressure ($p_0 < p_c$). 
Maintaining constant confining stresses and pore pressure, the axial loading was increased ($\sigma_1 > \sigma_2 = \sigma_3 $) until the specimen failed. 
The material properties and model parameters are listed in Table~\ref{tab: properties}, where $\upsilon$ denotes Poisson's ratio.

\begin{table}[h!]
    \centering
    \caption{Parameters used in the proposed model}
    \label{tab: properties}
    \begin{tabular}{cccccccccc}
    \toprule
        $E$ [GPa]  & $\upsilon$ & $G_c$ [N/m] &  $f_t$ [MPa] & $\alpha_0$ & A & $\kappa$ & $l_\mathrm{t}/l_\mathrm{c}$ & $\sigma_\mathrm{inf}$ [MPa] & $\sigma_\mathrm{sup}$ [MPa]\\
        \midrule
        19 & 0.12 & 60 & 4.0 & 0.8 & 1.2 & 1 & 39 & -100 & 0\\
        \bottomrule
    \end{tabular}\\
\end{table}

\begin{figure}[h!]
    \centering
    \includegraphics[width=0.75\linewidth]{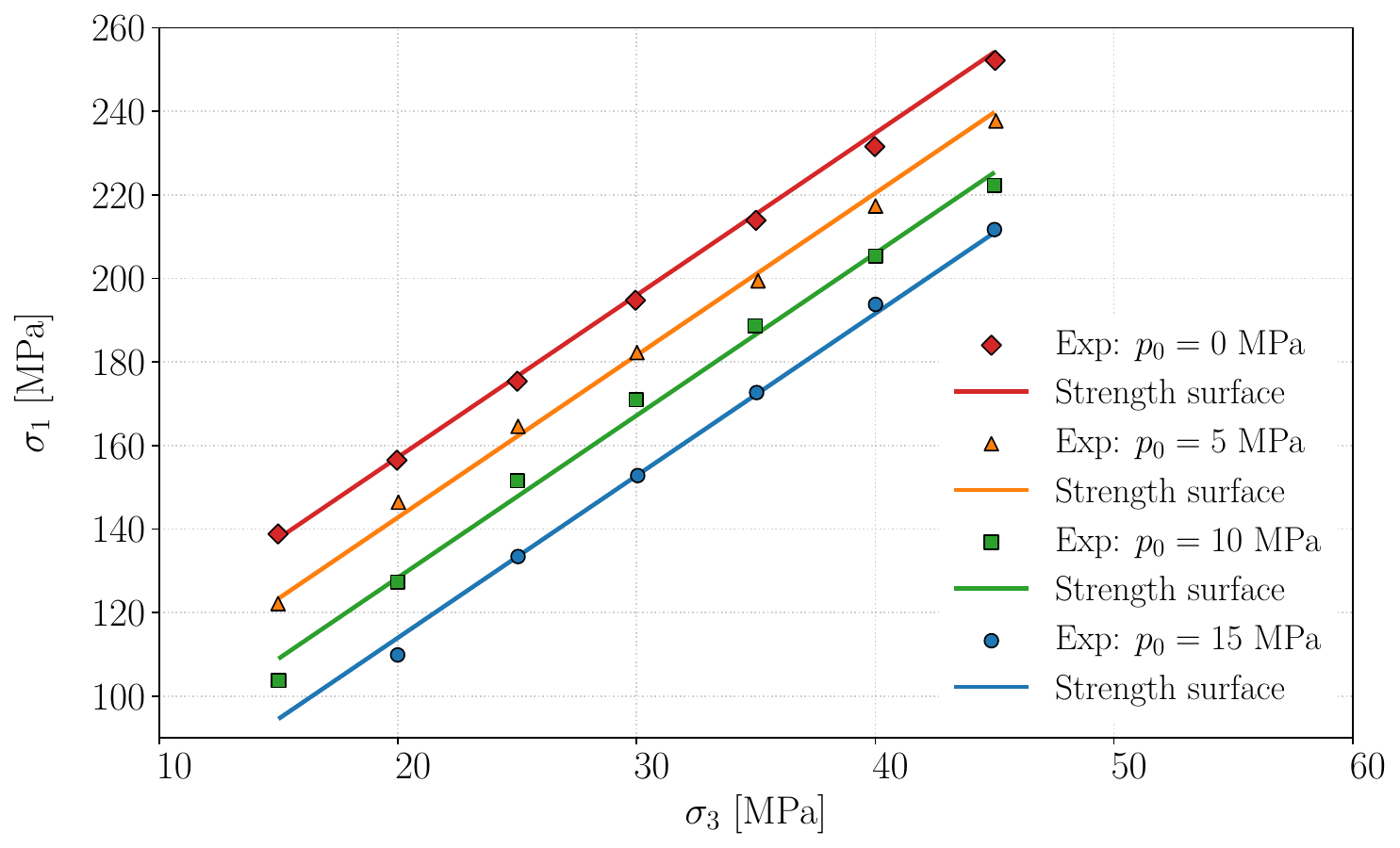}
    \caption{Comparisons of strength surfaces against the experimental results of gray sandstone under triaxial compression with different pore pressures ($p_0$=0, 5, 10, and 15~MPa). The data are extracted from~\citet{ZHU2023103789}.}
    \label{fig:comparison}
\end{figure}

Fig.~\ref{fig:comparison} plots the reported peak axial stresses $\sigma_1$, which are considered to represent the onset of fracturing, and the corresponding confining stress $\sigma_3$ under a given pore pressure $p_0$, together with our proposed strength surfaces. 
At a given pore pressure, the peak axial stress varies linearly with the confining stress, and this linear relation translates with changes in the pore pressure.
The proposed model successfully captures both the linear relation between $\sigma_1$ and $\sigma_3$ and the translation of the strength surfaces.

\subsection{Comparison against the mixed formulation}

On the other hand, if we base the strength surface for open microcracks on the mixed phase-field evolution formulation (Eq.~\eqref{eq: uncoupled phase-field evolution}), where the fluid pressure $p$ is taken as the independent variable rather than the fluid content $\xi$, we would have the following two different strength surfaces for open microcracks:

\begin{equation}
\label{eq:ss_open_coupled_uncoupled}
    \begin{aligned}
        \mathcal{F}_\mathrm{open}(\bm{\sigma},d_0) =
        \begin{dcases}
            \frac{\| \bm{\sigma}_\mathrm{dev} \|^2}{2\mu} + \frac{(\sigma_\mathrm{sph} + p)^2}{K} - \frac{G_c}{l_\mathrm{ch}}  &\text{Consistent formulation}\\
            \frac{\|\bm{\sigma}_{\mathrm{dev}}\|^{2}}{2\mu} + \frac{(\sigma_\mathrm{sph} + \alpha_0 p)^{2}}{K} - \frac{G_{c}}{l_\mathrm{ch}} - p^2 \frac{(1 - \alpha_0)^2}{K} &\text{Mixed formulation}
            .
        \end{dcases}
    \end{aligned}
\end{equation}

\begin{figure}[h!]
    \centering
    \begin{subfigure}{0.49\textwidth}
        \centering
        \includegraphics[width=\textwidth]{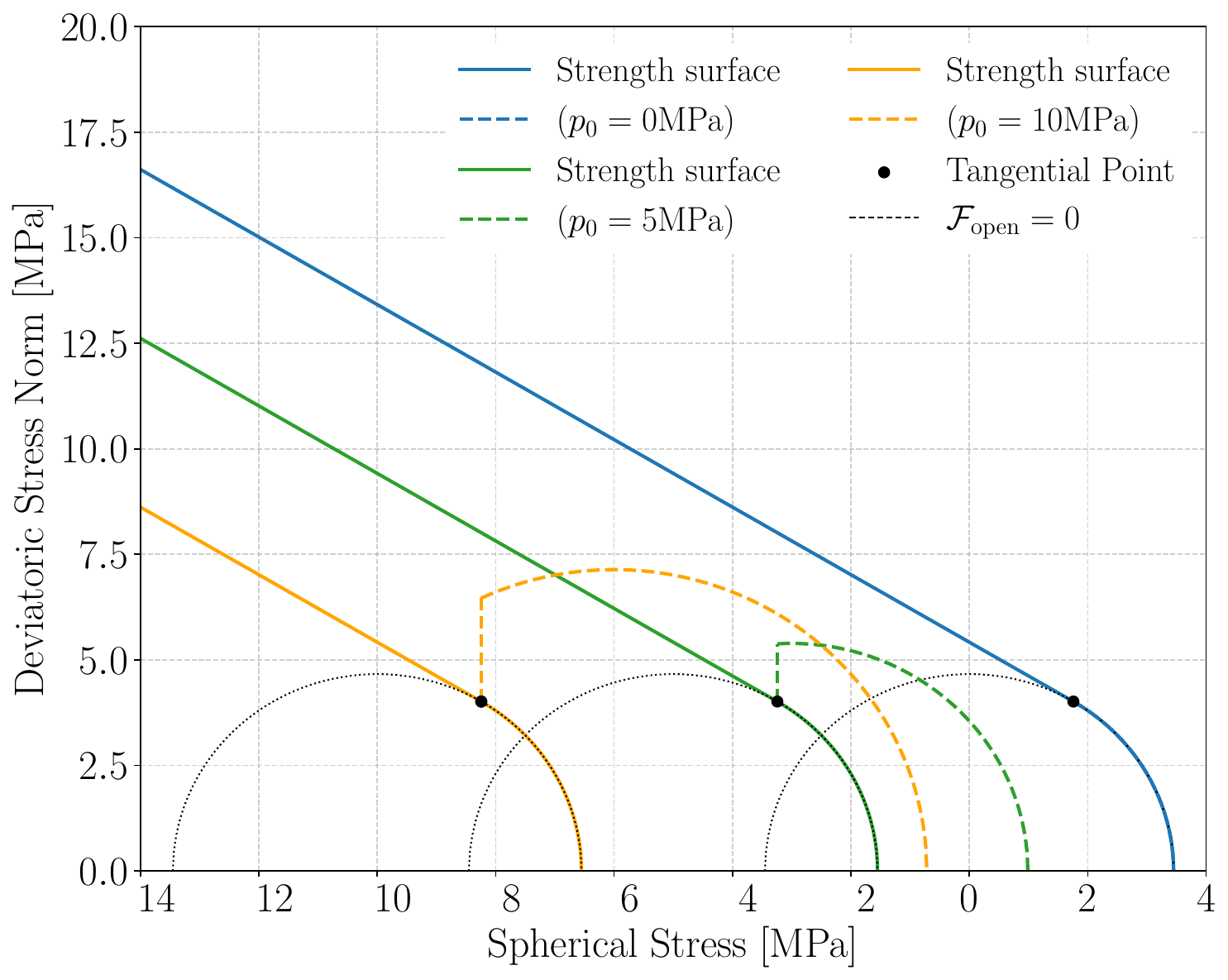}
        \caption{}
    \end{subfigure}
    \hfill 
    \begin{subfigure}{0.49\textwidth}
        \centering
        \includegraphics[width=\textwidth]{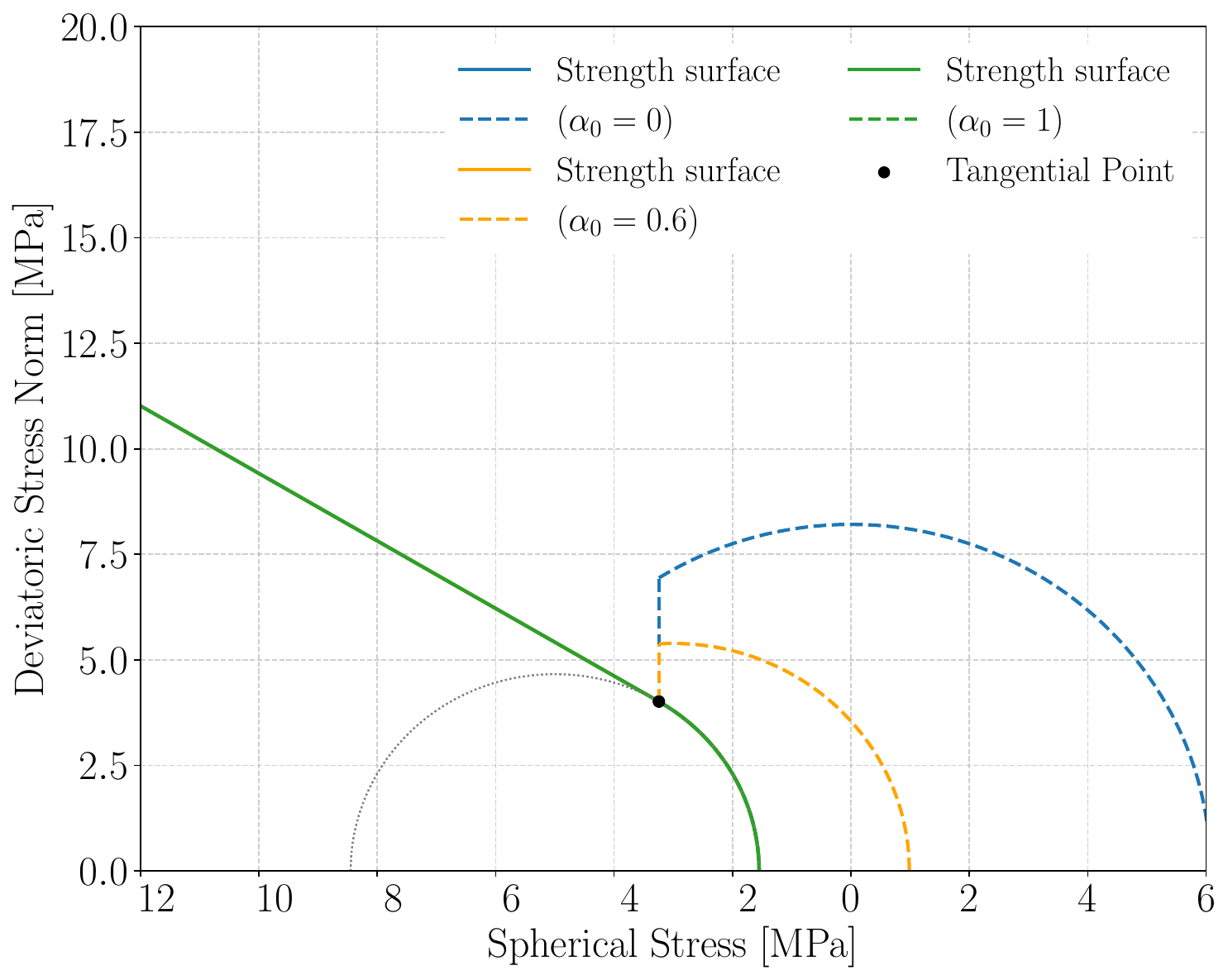}
        \caption{}
    \end{subfigure}
    \caption{Comparisons of strength surfaces (a) with different pore pressures ($p$=0, 5, and 10~MPa) under $\alpha_0=0.6$, and (b) with different Biot's coefficients ($\alpha$=0, 0.6 and 1) under $p=5$~MPa. The strength surfaces are plotted in ($\sigma_\mathrm{sph}$, $\| \bm{\sigma}_\mathrm{dev} \|$) space. The solid lines denote the strength surfaces of the consistent formulation, while the dashed lines denote the strength surfaces of the mixed formulation. The black dotted lines are the strength surfaces for tensile fracture derived from the consistent formulation. }
    \label{fig:ss_comparison}
\end{figure}

Note that the phase-field driving force for closed microcracks remains unaffected (Eq.~\eqref{eq: evolution of phase field}-2)\footnote{If the strength surfaces are derived from the Gibbs free energy by treating $\bm{\sigma}$, $p$, $\epsp$, and $d$ as independent variables, this issue of mixed partial differentiation can be avoided (see~\ref{app:gibbs}).}. 
We plot the complete strength surfaces under different pore pressures ($p$=0, 5, and 10~MPa) with $\alpha_0=0.6$ in Fig.~\ref{fig:ss_comparison}a, and under different Biot coefficients ($\alpha_0$=0, 0.6 and 1) with $p=5$~MPa in Fig.~\ref{fig:ss_comparison}b using the parameters listed in Table~\ref{tab: parameter_comparison}.
Here, variations in the characteristic length $l_\mathrm{ch}$ are not considered, and accordingly, extreme values for $\sigma_\mathrm{inf}$ and $\sigma_\mathrm{sup}$ are assigned.
Also, to highlight the implications of the mixed formulation, the mixed formulation in Eq.~\eqref{eq:ss_open_coupled_uncoupled}-2 is plotted together with the consistent surfaces.

\begin{table}[h!]
    \centering
    \caption{Parameters used for comparison}
    \label{tab: parameter_comparison}
    \begin{tabular}{cccccccccc}
    \toprule
        $E$ [GPa]  & $\upsilon$ & $G_c$ [N/m] &  $f_t$ [MPa] & $\alpha_0$ & A & $\kappa$ & $l_\mathrm{t}/l_\mathrm{c}$ & $\sigma_\mathrm{inf}$ [MPa] & $\sigma_\mathrm{sup}$ [MPa]\\
        \midrule
        100 & 0.15 & 120 & 5.0 & 0.6 & 0.8 & 1 & 1 & -1000 & -900\\
        \bottomrule
    \end{tabular}\\
\end{table}

For the dry case ($p=0$), the consistent and mixed formulations are identical.
However, for saturated porous media ($p>0$), the strength surfaces derived from the mixed formulation exhibit a discontinuity at the microcrack opening-closure transition, i.e., $\mathrm{tr}\left[ \bm{s}^\mathrm{p} \right] = 0$ (Fig.~\ref{fig:ss_comparison}a).
Under the consistent formulation, the pore pressure solely induces a global translation of the strength surface along the spherical stress axis~\citep{ZHU2023103789}. 
Furthermore, for $\alpha_0 = 1$, both the consistent and mixed formulations coincide; conversely, for $\alpha_0 < 1$, the mixed formulation again induces a discontinuity at the microcrack opening-closure transition (Fig.~\ref{fig:ss_comparison}b).

\begin{figure}[ht!]
    \centering
    \begin{subfigure}{0.47\textwidth}
        \centering
        \includegraphics[width=\textwidth]{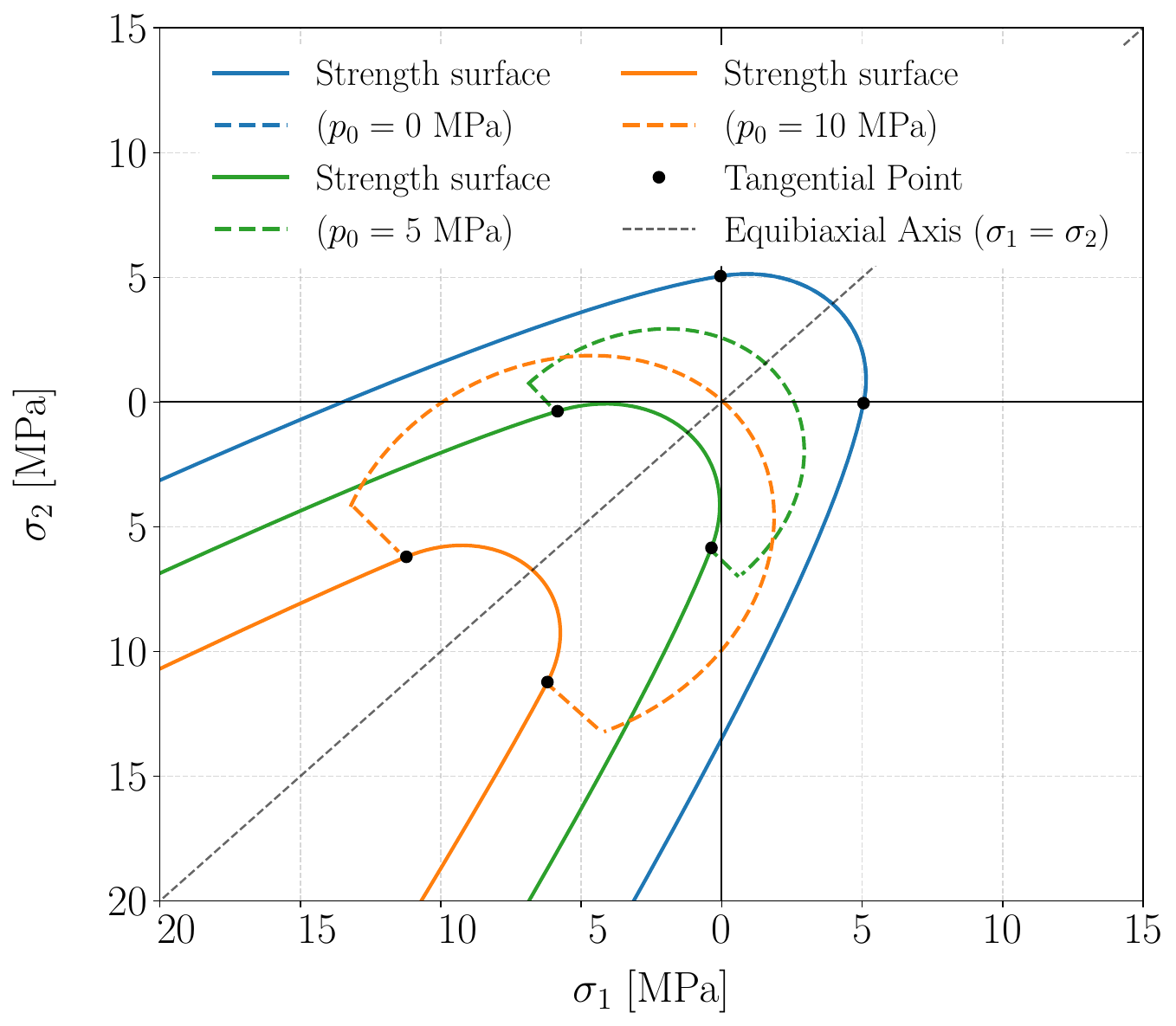}
        \caption{}
    \end{subfigure}
    \hfill 
    \begin{subfigure}{0.51\textwidth}
        \centering
        \includegraphics[width=\textwidth]{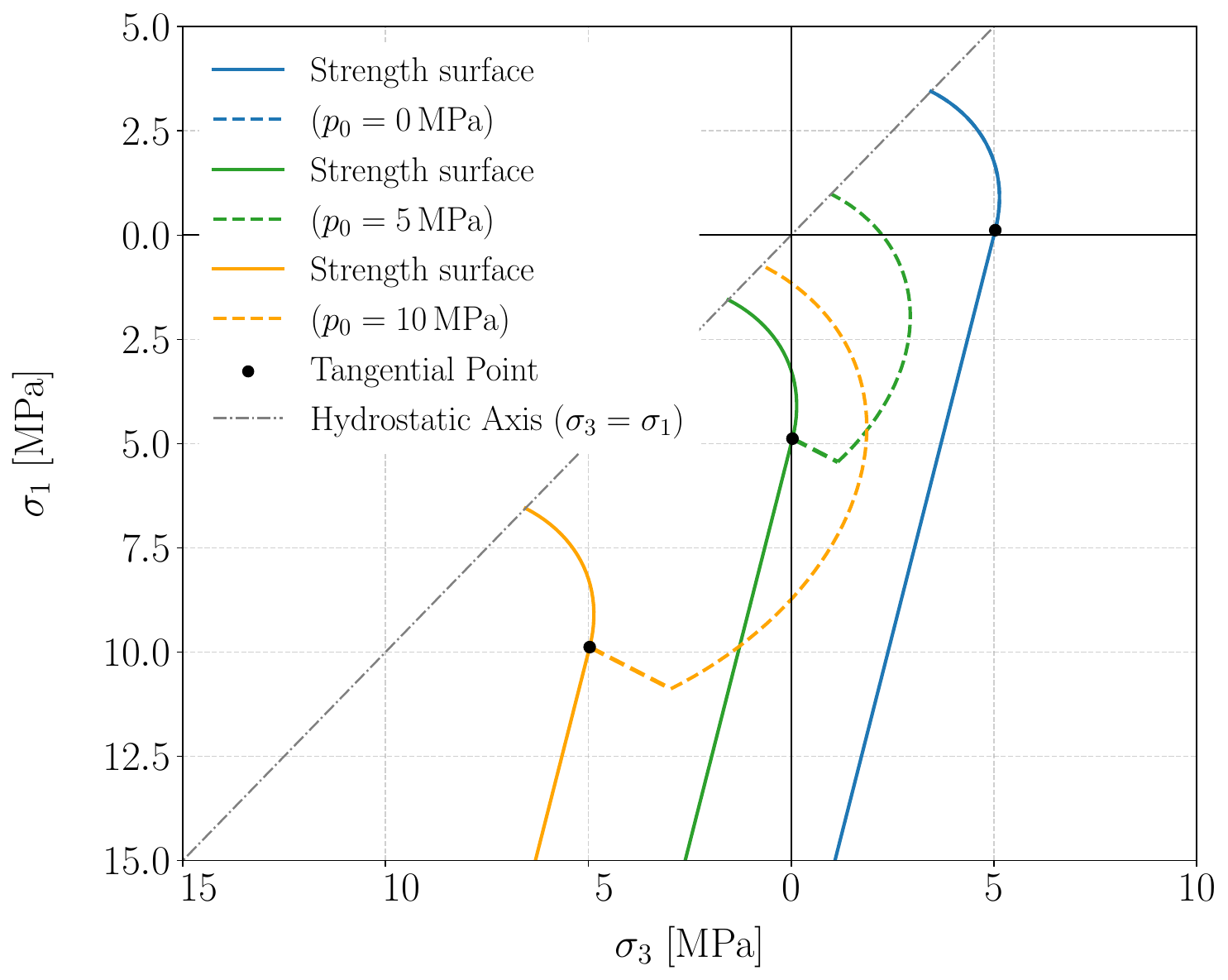}
        \caption{}
    \end{subfigure}
    \caption{Comparisons of strength surfaces under (a) biaxial loading condition with the assumption of plane-strain, and (b) triaxial extension condition plotted in principal stress space with different pore pressures ($p$=0, 5, and 10~MPa). The solid lines denote the strength surfaces of the consistent formulation, while the dashed lines denote the strength surfaces of the mixed formulation.}
    \label{fig:strength_comparison}
\end{figure}

On the other hand, using the consistent formulation, the strength surface for tensile fracture is tangent to that for compressive-shear fracture, resulting in a smooth transition between the tensile fracture and compressive-shear fracture modes.
To highlight the discontinuous yield surface produced by the mixed formulation, the proposed strength surfaces are plotted for biaxial loading in a plane strain condition (Fig.~\ref{fig:strength_comparison}a) and for triaxial extension (Fig.~\ref{fig:strength_comparison}b) with different pore pressures ($p$=0, 5, and 10~MPa) in the principal stress space.


\subsection{Impacts of the individual degradation and plastic flow rule}
\label{sec:diff flow rule}

In \cite{ULLOA2022104684}, the volumetric and deviatoric parts of the stiffness tensor are degraded individually using different functions, and a non-associative flow rule is applied for the plastic flow.
This section explores the impacts of these treatments on the strength surfaces.


\subsubsection{Individual degradation functions}
With the Mori-Tanaka homogeneity scheme~\citep{MORI1973571}, one can degrade the volumetric and deviatoric parts of the stiffness tensors individually as follows~\citep{ULLOA2022115084}
\begin{equation}
\label{eq:respective degradation}
    \begin{aligned}
        \mathbb{C}_\mathrm{dam}(d) &= 3Kg_K(d)\mathbb{J} + 2\mu g_\mu(d)\mathbb{K}, \\
        \mathbb{H}(d) &= 3Kg_{p,K}(d)\mathbb{J} + 2\mu g_{p,\mu}(d)\mathbb{K}
    \end{aligned}
\end{equation}
where $g_K(d)$ denotes the degradation function for the volumetric part and $g_\mu(d)$ denotes the degradation function for the deviatoric.
Several forms of $g_K(d)$ have been proposed to better describe the specific behaviors of materials~\citep{pham2011gradient, bourdin2000numerical, kuhn2015degradation, karma2001phase, WU201772}. 
In this paper, we take $g_K(d) = g(d)$, which yields
\begin{equation}
\label{eq:g_mu}
    g_\mu(d) = \frac{g(d)}{g(d) + \frac{b_\mu}{b_K}[1 - g(d)]} \quad \mathrm{with} \quad \frac{b_\mu}{b_K} = \frac{2(5 - \upsilon)(1 - 2\upsilon)}{5(2 - \upsilon)(1 + \upsilon)} 
    .
\end{equation}
For the hardening tensor, the degradation functions read
\begin{equation}
\label{eq:g_p}
    g_{p,K} = \frac{g(d)}{1 - g(d)}, \quad g_{p,\mu} = \frac{g_\mu(d)}{1 - g_\mu(d)}
\end{equation}

Using these functions that degrade the volumetric and deviatoric parts of the stiffness individually, we can obtain the strength surfaces as
\begin{equation}
\label{eq:vd}
    \begin{aligned}
        \mathcal{F}^\mathrm{VD}_\mathrm{open}(\bm{\sigma},d_0) =& \frac{b_\mu}{b_K}\frac{\| \bm{\sigma}_\mathrm{dev} \|^2}{2\mu} + \frac{(\sigma_\mathrm{sph} + p)^2}{K} - \frac{G_c}{l_\mathrm{ch}} = 0 \\
        \mathcal{F}^\mathrm{VD}_\mathrm{close}(\bm{\sigma}, d_0) =& \| \bm{\sigma}_\mathrm{dev} \| + A(\sigma_\mathrm{sph} + p) - \sqrt{\frac{G_c \chi_1}{l_\mathrm{ch}}} = 0
    \end{aligned}
\end{equation}
where $\chi_1 = A^2K + 2\mu \frac{b_K}{b_\mu}$ and superscript ``VD'' denotes the individual degradation of the volumetric and deviatoric parts.

\subsubsection{Non-associative flow rule}
When a non-associative flow rule is applied, the plastic potential is replaced with:
\begin{equation}
    h^\mathrm{p}(\lsp) := \| \lspd \| + \frac{A_\theta}{3}\mathrm{tr}[\lsp]
\end{equation}
where $A_\theta$ is the dilation coefficient, and commonly $A > A_\theta > 0$ for geomaterials. 
The plastic flow direction reads
\begin{equation}
    \mathbf{D} = \mathbf{V} + \frac{A_\theta}{3}\mathbf{I}
    .
\end{equation}
Accordingly the strength surface for the closed case ($ \mathcal{F}^\mathrm{NA}_\mathrm{close}$) becomes
\begin{equation}
\label{eq:nafr}
    \begin{aligned}
        \mathcal{F}^\mathrm{NAFR}_\mathrm{close}(\bm{\sigma}, d_0) =& \| \bm{\sigma}_\mathrm{dev} \| + A(\sigma_\mathrm{sph} + p) - \left(2\mu + A A_\theta K \right)\sqrt{\frac{G_c}{l_\mathrm{ch}\chi_2}} = 0
    \end{aligned}
\end{equation}
where $\chi_2 = A_\theta^2K + 2\mu$.
Note that the associated flow rule can be viewed as a special case.
With $A = A_\theta$, Eq.~\eqref{eq:nafr} reduces to Eq.~\eqref{eq:ss_close}. 

\begin{figure}[h!]
    \centering
    \includegraphics[width=0.6\linewidth]{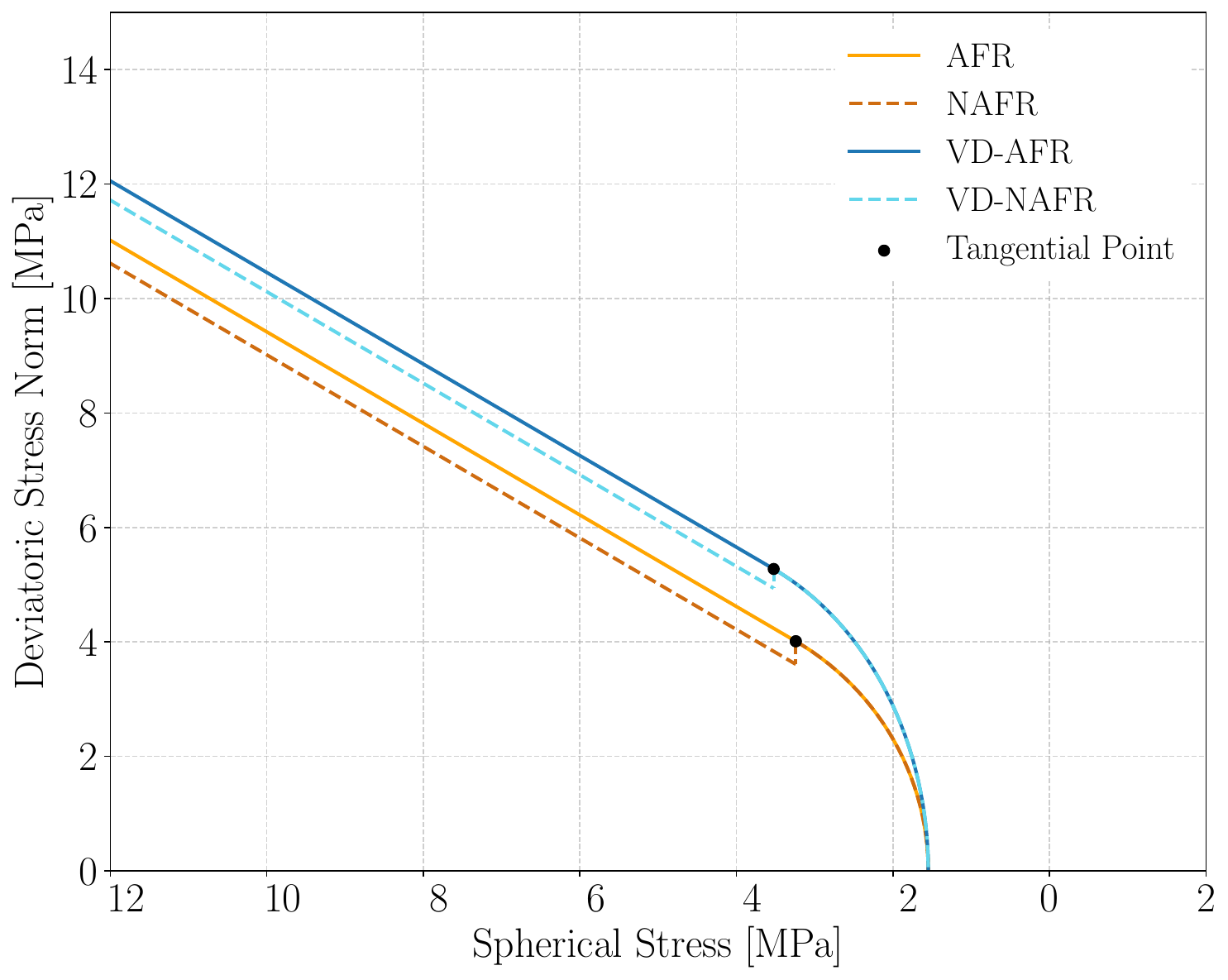}
    \caption{Comparisons of strength surfaces considering decomposed fourth-order tensors and plastic flow rules. VD denotes the respective degradation on the volumetric and deviatoric parts of the fourth-order tensors. AFR is the abbreviation for associative flow rule, and NAFR is the abbreviation for non-associative flow rule. The solid lines represent the strength surfaces with the associative flow rule, and the dashed lines represent the strength surfaces with the non-associative flow rule.}
    \label{fig:diff_flow_rule}
\end{figure}

Fig.~\ref{fig:diff_flow_rule} plots the four different strength surfaces: 1) associative flow rule (AFR), 2) associative flow rule with individual degradation for the volumetric and deviatoric parts (VD-ARF), 3) non-associative flow rule (NAFR), and 4) non-associative flow rule with individual degradation for the volumetric and deviatoric parts (VD-NAFR).
The initial pore pressure is $p_0 = 5$ MPa, and the dilation coefficient is set to $A_\theta = 0.25$.
The remaining parameters are listed in Table~\ref{tab: parameter_comparison}.

In the absence of deviatoric stress ($\| \bm{\sigma}_\mathrm{dev} \| = 0$), all strength surfaces coincide. 
With increasing deviatoric stress, the strength surfaces with and without individual degradation deviate from one another but remain continuous up to the microcrack opening-closure transition. 
At this transition point, the strength surfaces derived using the non-associative flow rule become discontinuous. 
Due to the dilation coefficient $A_\theta$, the strength decreases significantly for the closed microcracks. 

In the following numerical experiments, we employ an associative flow rule ($A = A_\theta$) to ensure the continuity in the strength surfaces.
Although the use of the non-associative flow rule is common for geomaterials, its necessity has been questioned by~\cite{zhu2016analytical,marigo2019micromechanical}.


\section{Numerical implementation}
\label{sec: numerical implementation}

\subsection{Weak form of the governing equations}
The weak forms of the governing equations for plastic evolution, phase-field evolution, and mass balance are given as:
\begin{equation}
\label{eq: ge}
    \begin{aligned}
        &\int_{\Omega}\nabla\bm{\omega}_u \cdot \bm{\sigma} \mathrm{d}V - \int_{\Omega}\bm{b}\cdot\bm{\omega}_u \mathrm{d}V -  \int_{\partial\Omega_t}\bar{\bm{t}}\cdot\bm{\omega}_u\mathrm{d}S = 0  \\
        &\int_{\Omega} \omega_d (-e^d) \mathrm{d}V=0 \\
        &\int_\Omega \omega_p \frac{\partial \xi(\bm{\varepsilon},d)}{\partial t} \mathrm{d}V + \int_\Omega \frac{\bm{K}}{\nu}\nabla p \cdot \nabla \omega_p \mathrm{d}V = \int_\Omega \omega_p Q \mathrm{d}V - \int_{\partial \Omega_q} \omega_p \bar{\bm{q}} \mathrm{d}S
    \end{aligned}
\end{equation}
where $\bm{\omega}_u$, $\omega_d$, and $\omega_p$ are the test functions, $\bm{b}$ denotes the body force, and $\bar{\bm{t}}$ is the traction applied to the surface of the domain.

The expanded forms of Eq.~\eqref{eq: ge} read
\begin{equation}
\label{eq:microforce-equal}
    \begin{dcases}
        \int_{\Omega}\nabla\bm{\omega}_u \cdot \left( \mathbb{C}_\mathrm{dam}(d):\bm{\varepsilon} - \alpha p \mathbf{I} \right) \mathrm{d}V - \int_{\Omega}\bm{b}\cdot\bm{\omega}_u \mathrm{d}V -  \int_{\partial\Omega_t}\bar{\bm{t}}\cdot\bm{\omega}_u\mathrm{d}S = 0 \quad &\text{if} \quad \mathrm{tr}[\bm{s}^\mathrm{p}] = 0 \\
        \int_{\Omega}\nabla\bm{\omega}_u \cdot \left( \mathbb{C}:(\bm{\varepsilon} - \epsp) - \alpha_0 p \mathbf{I} \right) \mathrm{d}V - \int_{\Omega}\bm{b}\cdot\bm{\omega}_u \mathrm{d}V -  \int_{\partial\Omega_t}\bar{\bm{t}}\cdot\bm{\omega}_u\mathrm{d}S = 0 \quad &\text{if} \quad \mathrm{tr}[\bm{s}^\mathrm{p}] < 0
    \end{dcases}
\end{equation}
\begin{equation}
    \begin{dcases}
        \int_{\Omega} \omega_d \frac{1}{2}\frac{\partial g(d)}{\partial d} \left[ \bm{\varepsilon}:\mathbb{C}:\bm{\varepsilon} + \frac{(1 - \alpha_0)^2}{K}p^2 + 2(1 - \alpha_0)p\mathrm{tr}[\bm{\varepsilon}] \right] \mathrm{d}V \\
        \qquad - \int_{\Omega} \omega_d \frac{2G_c d}{\pi \ell} \mathrm{d}V + \int_\Omega \frac{2G_c \ell}{\pi}\nabla\omega_d \cdot \nabla d \mathrm{d}V =0  \quad \text{if} \quad \mathrm{tr}[\bm{s}^\mathrm{p}] = 0  \\
        \int_{\Omega} \omega_d \frac{1}{2}\frac{\partial g_p(d)}{\partial d} \epsp :\mathbb{C}:\epsp \mathrm{d}V \\
        \qquad - \int_{\Omega} \omega_d \frac{2G_c d}{\pi \ell} \mathrm{d}V + \int_\Omega \frac{2G_c \ell}{\pi}\nabla\omega_d \cdot \nabla d \mathrm{d}V =0  \quad \text{if} \quad \mathrm{tr}[\bm{s}^\mathrm{p}] < 0 
    \end{dcases}
\end{equation}
\begin{equation}
\label{eq: govern fluid}
    \begin{dcases}
        \int_{\Omega} \omega_p \left( \alpha \frac{\partial \mathrm{tr}[\bm{\varepsilon}]}{\partial t}\mathrm{tr}[\bm{\varepsilon}] + \frac{1}{M}\frac{\partial p}{\partial t} \right) \mathrm{d}V \\
        \qquad + \int_\Omega \frac{\bm{K}}{\nu}\nabla p \cdot \nabla \omega_p \mathrm{d}V - \int_\Omega \omega_p Q \mathrm{d}V + \int_{\partial \Omega_q} \omega_p \bar{\bm{q}} \mathrm{d}S = 0  \quad \text{if} \quad \mathrm{tr}[\bm{s}^\mathrm{p}] = 0  \\
        \int_{\Omega} \omega_p \left( \alpha_0 \frac{\partial \mathrm{tr}[\bm{\varepsilon} - \bm{\varepsilon}^\mathrm{p}]}{\partial t} + \frac{1}{M_0} \frac{\partial p}{\partial t} + \frac{\partial \mathrm{tr}[\bm{\varepsilon}^\mathrm{p}]}{\partial t} \right) \mathrm{d}V \\
        \qquad + \int_\Omega \frac{\bm{K}}{\nu}\nabla p \cdot \nabla \omega_p \mathrm{d}V - \int_\Omega \omega_p Q \mathrm{d}V + \int_{\partial \Omega_q} \omega_p \bar{\bm{q}} \mathrm{d}S = 0  \quad \text{if} \quad \mathrm{tr}[\bm{s}^\mathrm{p}] < 0 
    \end{dcases}
\end{equation}
where the temporal and spatial discretizations are provided in~\ref{Temporal and spatial discretizations of the governing equations}.

The local return-mapping algorithm~\citep{simo1998computational, ComputationalMethodsforPlasticity, Borja2013} is employed to evaluate the update of the elastoplastic variables, e.g., plastic strain $\epsp$ and local stress $\bm{s}^\mathrm{p}$, under different stress states. 
For the dry case ($p = 0$), the procedure is identical to that presented in~\citet{li2025cohesive}. 
For $p \neq 0$, we can use Eq.~\ref{eq:trsp} for the trial stresses.

\subsection{Solution scheme}
The phase-field $d$ and the displacement $\bm{u}$ fields are normally solved in a staggered scheme, taking advantage of their respective convexity. 
On the other hand, for the coupled deformation ($\bm{u}$) and pore pressure $p$ system, a monolithic scheme is well established. 

Thus, to combine these systems, we employ a hybrid monolithic-staggered scheme to solve the three-field ($d$-$p$-$\bm{u}$) problem, where the $p$-$\bm{u}$ coupling is solved directly using a monolithic scheme, while iterations between $d$ and $p$-$\bm{u}$ are solved in a staggered scheme, as presented in Algorithm~\ref{alg:stag}.

\begin{algorithm}[h!]
	\renewcommand{\algorithmicrequire}{\textbf{Input:}}
	\renewcommand{\algorithmicensure}{\textbf{Output:}}
	\caption{The hybrid monolithic-staggered scheme for solving coupled $d - p - \bm{u}$ problem }
    \label{alg:stag}
	\begin{algorithmic}[1]
		\STATE \texttt{/* For $n$ time step :}
        \STATE Given primary fields at previous time step $d_{n-1}$, $p_{n-1}$, and $\bm{u}_{n-1}$ 
		\STATE Initiate iteration $k = 0$, $d_{n}^{k} = d_{n-1}$, $p_{n}^{k} = p_{n-1}$, and $\bm{u}_{n}^{k} = \bm{u}_{n-1}$  
		\WHILE{$err > tol$ \textbf{and} $k < max\_iter $ }
		\STATE $k \leftarrow k + 1$;
        \STATE Calculate $l_\mathrm{ch}^k$ with $\sigma_{\mathrm{min},n}^{k-1}$ from $\bm{u}_n^{k-1}$
        \STATE Solve the phase-field problem for $d_{n}^{k}$ \\ 
            \qquad using $d_{n}^{k-1}$, $p_{n}^{k-1}$, $\bm{u}_{n}^{k-1}$, and $l_\mathrm{ch}^k$; 
		\STATE Solve the displacement-pressure problem for $p_{n}^{k}$ and $\bm{u}_{n}^{k}$ \\ 
            \qquad using $d_{n}^{k}$, $p_{n}^{k-1}$, $\bm{u}_{n}^{k-1}$, and $l_\mathrm{ch}^k$;
		\STATE Check for convergence: $err = \mathrm{max}(\frac{\|d_{n}^{k} - d_{n}^{k-1}\|}{\|d_{n}^{k}\|},\frac{\|p_{n}^{k} - p_{n}^{k-1}\|}{\|p_{n}^{k}\|} , \frac{\|\bm{u}_{n}^{k} - \bm{u}_{n}^{k-1}\|}{\|\bm{u}_{n}^{k}\|})$.
		\ENDWHILE
		\STATE   Update $d_{n} = d_n^{k}$, $p_{n} = p_n^{k}$, and $\bm{u}_{n} = \bm{u}_n^{k}$.
	\end{algorithmic}  
\end{algorithm}

The respective Newton iterations read
\begin{equation}
\label{eq:d-newton}
    \begin{Bmatrix}
        \bm{d}
    \end{Bmatrix}_{n}
    =
    \begin{Bmatrix}
        \bm{d} 
    \end{Bmatrix}_{n-1}
    -
    \begin{bmatrix}
        \mathbf{K}^{dd}
    \end{bmatrix}^{-1}_{n-1}
    \begin{Bmatrix}
        \bm{r}^d
    \end{Bmatrix}_{n-1} \, ,
\end{equation}
\begin{equation}
\label{eq:p-u-monolithich-newton}
    \begin{Bmatrix}
        \bm{p} \\
        \bm{u} 
    \end{Bmatrix}_{n}
    =
    \begin{Bmatrix}
        \bm{p} \\
        \bm{u} 
    \end{Bmatrix}_{n-1}
    -
    \begin{bmatrix}
        \mathbf{K}^{pp} & \mathbf{K}^{p\bm{u}} \\
        \mathbf{K}^{\bm{u}p} & \mathbf{K}^{\bm{u}\bm{u}}
    \end{bmatrix}^{-1}_{n-1}
    \begin{Bmatrix}
        \bm{r}^p \\
        \bm{r}^{\bm{u}}
    \end{Bmatrix}_{n-1}
    ,
\end{equation}
where the residuals of each process and the Jacobian matrices are given in~\ref{Temporal and spatial discretizations of the governing equations}.

\section{Numerical examples}
\label{Numerical experiment}
This section presents numerical results obtained using the proposed phase-field framework. 
First, we consider a poroelastic fracture driven by fluid injection in the absence of plastic deformation. 
The results are compared with an analytical solution to demonstrate the impact of the poroelastic coupling term on solution accuracy. 
Second, we simulate a hydraulic fracturing example that includes plastic evolution to highlight the effects of plasticity. 
Finally, a biaxial compression test is simulated to reproduce compressive-shear fracture in a saturated quasi-brittle material.

Note that for elastoplastic computations, a small initial value of damage is required to avoid a numerical singularity in Eq.~\eqref{eq: gp}.
In the following cases, we set $d_0 = 0.999$ as the initial damage in the entire domain.

\subsection{Fluid driven fracturing in poro-elastic model}
\label{sec:verification_KGD}
A classical hydraulic fracturing benchmark case under plane strain conditions, known as the KGD (Kristianovich–Geertsma–de Klerk) model~\citep{kristianovitch1955formation, 10.2118/2458-PA}, was employed to evaluate the discrepancy between numerical simulation results obtained using the consistent and mixed formulations. 
The time-dependent analytical solutions of the KGD model formulated for fluid injection within an infinite plane have been derived by~\citet{garagash2006plane,santillan2017phase} (refer to~\ref{sec:KGD analytical solutions} for details). 

In their analytical solution, the dimensionless viscosity $\mathcal{M}$ is given as
\begin{equation}
\label{eq:dimensionless_mu}
    \mathcal{M} = \frac{\mu^\prime Q}{E^\prime}\left( \frac{E^\prime}{K^\prime} \right)^4
\end{equation}
where $\mu^\prime = 12 \mu$, $E^\prime = \frac{E}{1 - \upsilon^2}$, $K^\prime = \sqrt{\frac{32 G_c^\mathrm{eff} E^\prime}{\pi}}$. 
Note that for the critical energy release rate $G_c$, we use the ``effective'' value that accounts for the mesh size $h$ and the length scale parameter $\ell$~\citep{bourdin2008variational, tanne2018crack, yoshioka2020crack} given as 
\begin{equation}
    G_c^\mathrm{eff} = G_c \left( 1 + \frac{h}{4c_n\ell} \right)
    ,
\end{equation}
with $c_n = \frac{2}{3}$ for the AT1 model and $h / \ell = \frac{1}{2}$ in this case. 
Using the parameters listed in Table~\ref{tab:hf properties}, we obtain $\mathcal{M} = 3.8\times10^{-7}$. 
For $\mathcal{M} < \mathcal{M}_c = 3.4\times10^{-3}$, the hydraulic fracturing process falls within the toughness-dominated regime, where the impact of fracture toughness dominates over fluid viscous dissipation~\citep{garagash2006plane, santillan2017phase}.

\begin{table}[h!]
    \centering
    \caption{Properties of the matrix and fluid}
    \label{tab:hf properties}
    \begin{tabular}{p{5cm}cc}
    \toprule
        Properties & Value & Unit\\
        \midrule
        Young's modulus ($E$) & 17 & GPa \\
        Poisson's ratio ($\upsilon$) & 0.2 & -\\
        Critical energy release rate ($G_c$) & 300 & N/m\\
        Biot's coefficient ($\alpha_0$) & 0.0 & -\\
        Porosity ($\phi_0$) & 0.0 & - \\
        Fluid compressibility ($c_f$) & 0.0 & - \\
        Permeability ($K_\mathrm{m}$) & $1 \times 10^{-19}$ & $\text{m}^2$\\
        Fluid viscosity ($\mu$) & $1 \times 10^{-8}$ & Pa$\cdot$s \\
        Injection rate ($Q$) & $2 \times 10^{-3}$ & $\text{m}^2 / \text{s}$ \\
        \bottomrule
    \end{tabular}\\
\end{table}

\begin{figure}[h!]
    \centering
    \begin{subfigure}[b]{0.33\textwidth}
        \centering
        \includegraphics[width=\textwidth]{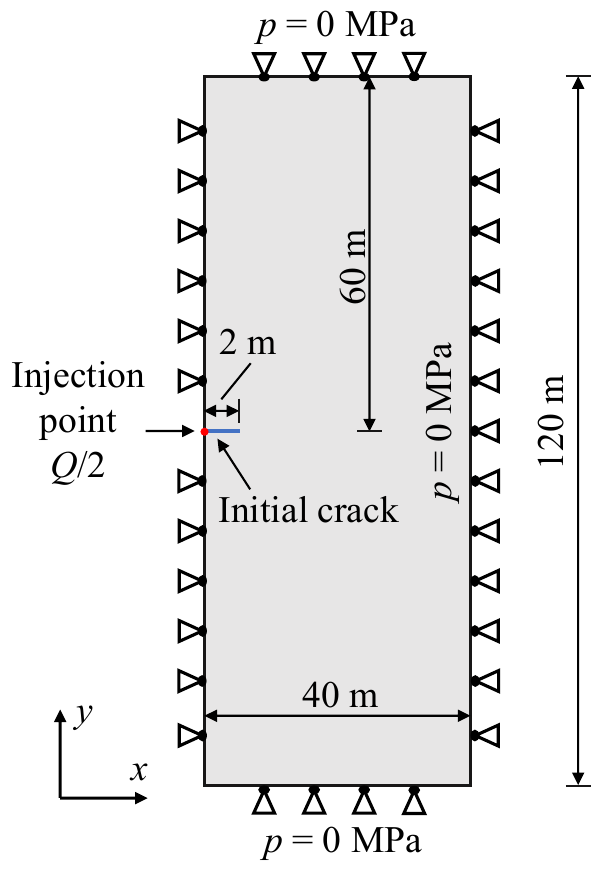}
        \caption{}
        \label{}
    \end{subfigure}
    \hfill
    \begin{subfigure}[b]{0.51\textwidth}
        \centering
        \includegraphics[width=\textwidth]{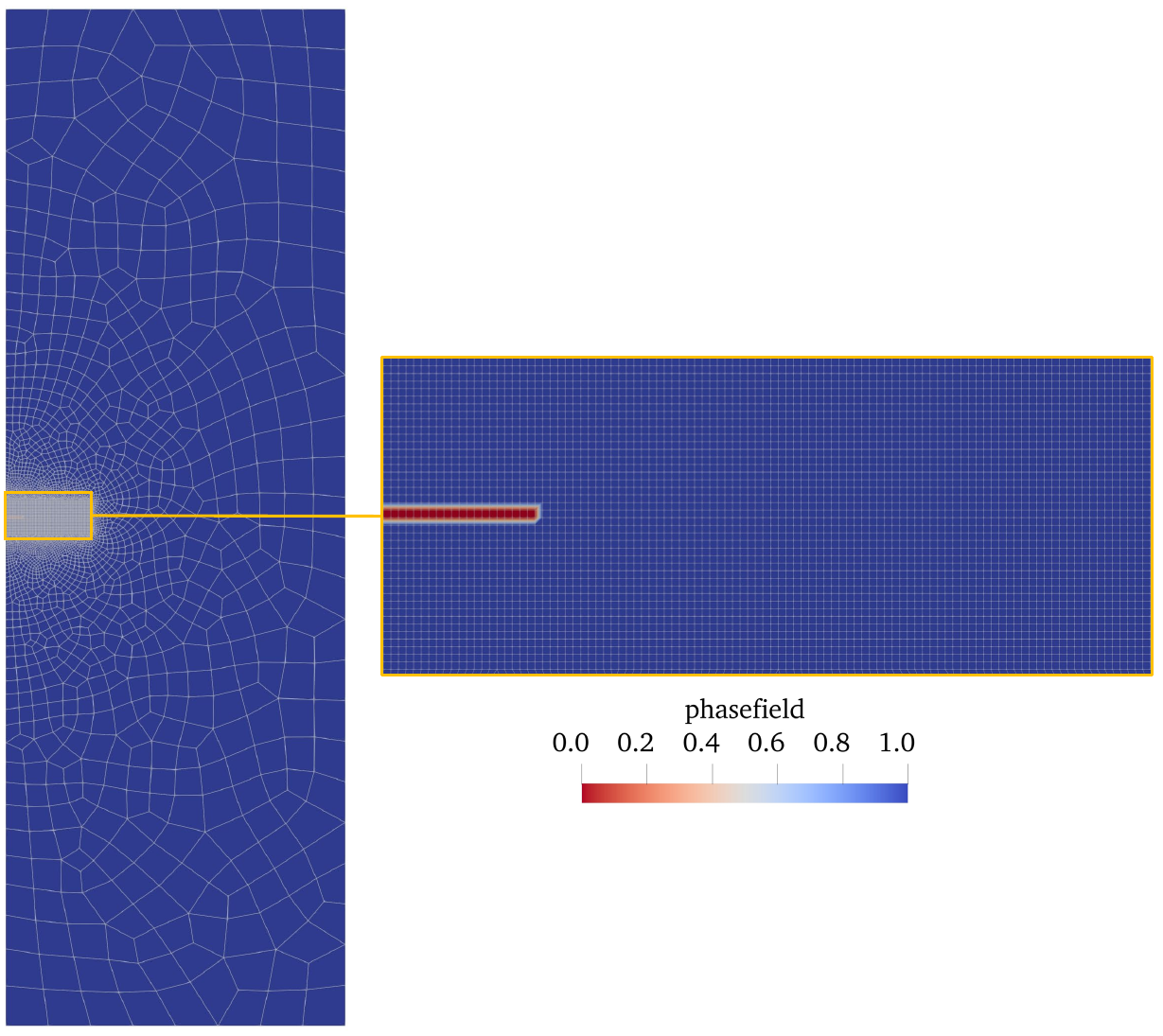}
        \caption{}
        \label{}
    \end{subfigure} 
    \caption{A schematic representation of (a) the geometry and (b) mesh of the KGD model.}
    \label{fig:KGD_model}
\end{figure}

Exploiting the symmetry of the infinite plane containing an internal crack, the numerical domain was simplified to a semi-infinite plane with a half-crack situated at the boundary, as illustrated in Fig.~\ref{fig:KGD_model}a.
Consequently, the injection rate in the numerical simulation was set to $Q_\mathrm{inj} = Q/2 = 1 \times 10^{-3} \, \text{m}^2/\text{s}$. 
To approximate an infinite domain, a sufficiently large rectangular domain of ([0m, 40m] $\times$ [0m, 120m]) was used, with an initial crack located at [0m, 2m] $\times$ [60m]. 
A locally refined mesh consisting of 7,254 quadrilateral elements was employed (Fig.~\ref{fig:KGD_model}b), with a minimum element size of $h = 0.1~\text{m}$. 
All four edges were constrained in their normal directions, and the pressure at the top, bottom, and right edges was set to $p = 0$.

\begin{figure}[h!]
    \centering
    \begin{subfigure}[b]{0.48\textwidth}
        \centering
        \includegraphics[width=\textwidth]{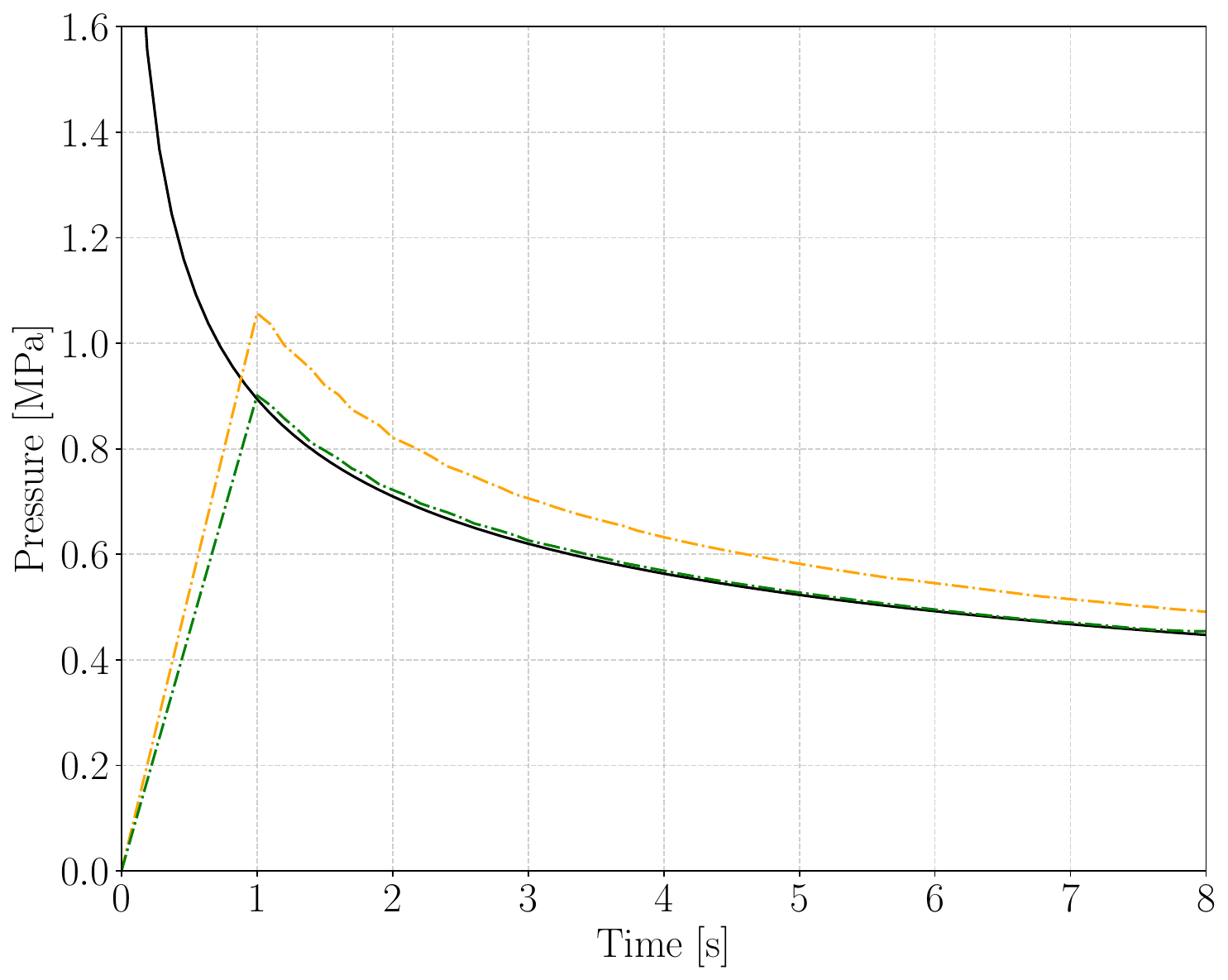}
        \caption{}
        \label{fig:KGD-pressure}
    \end{subfigure}
    \hfill
    \begin{subfigure}[b]{0.48\textwidth}
        \centering
        \includegraphics[width=\textwidth]{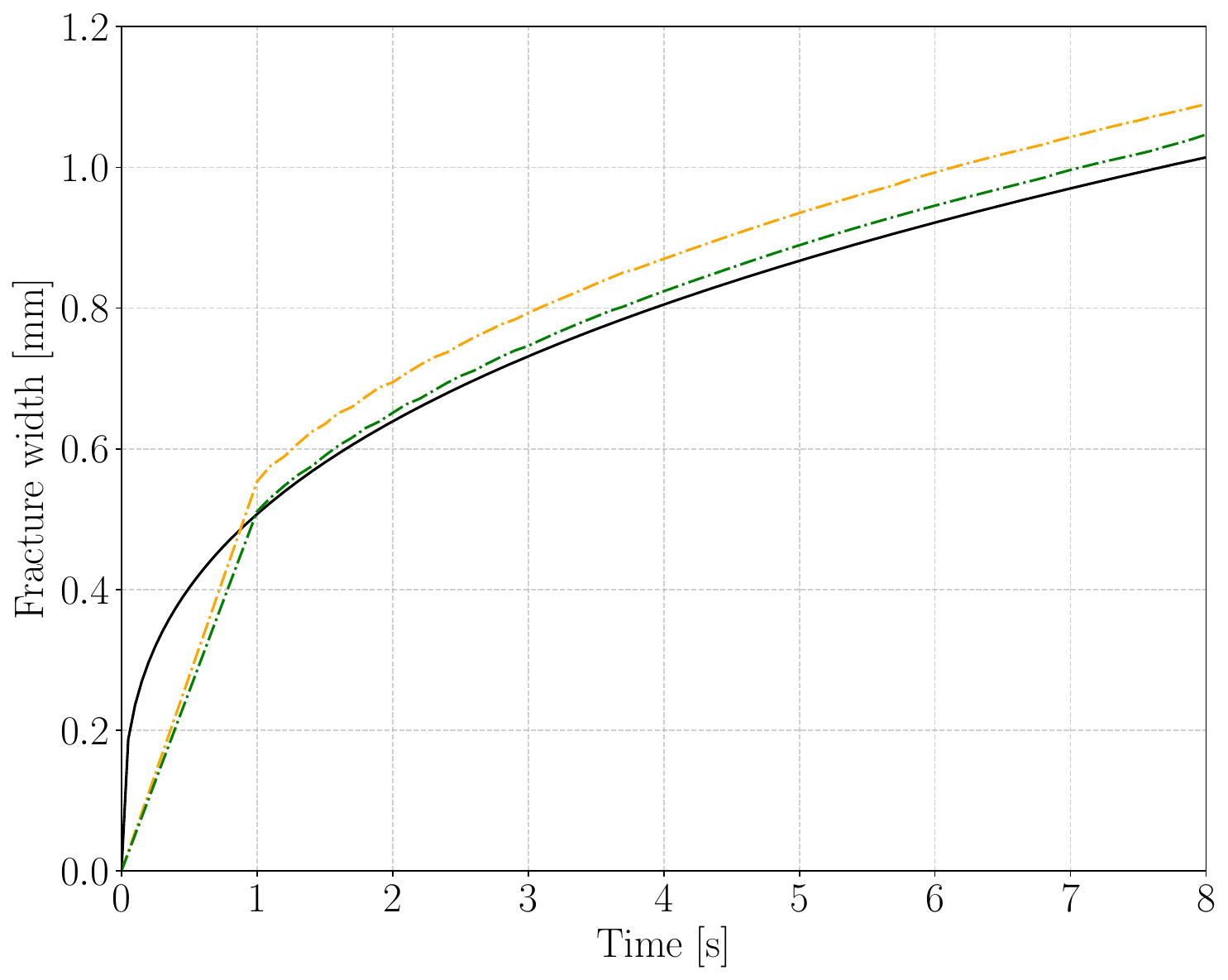}
        \caption{}
        \label{fig:KGD-width}
    \end{subfigure} \\
    \begin{subfigure}[b]{0.69\textwidth}
        \centering
        \includegraphics[width=\textwidth]{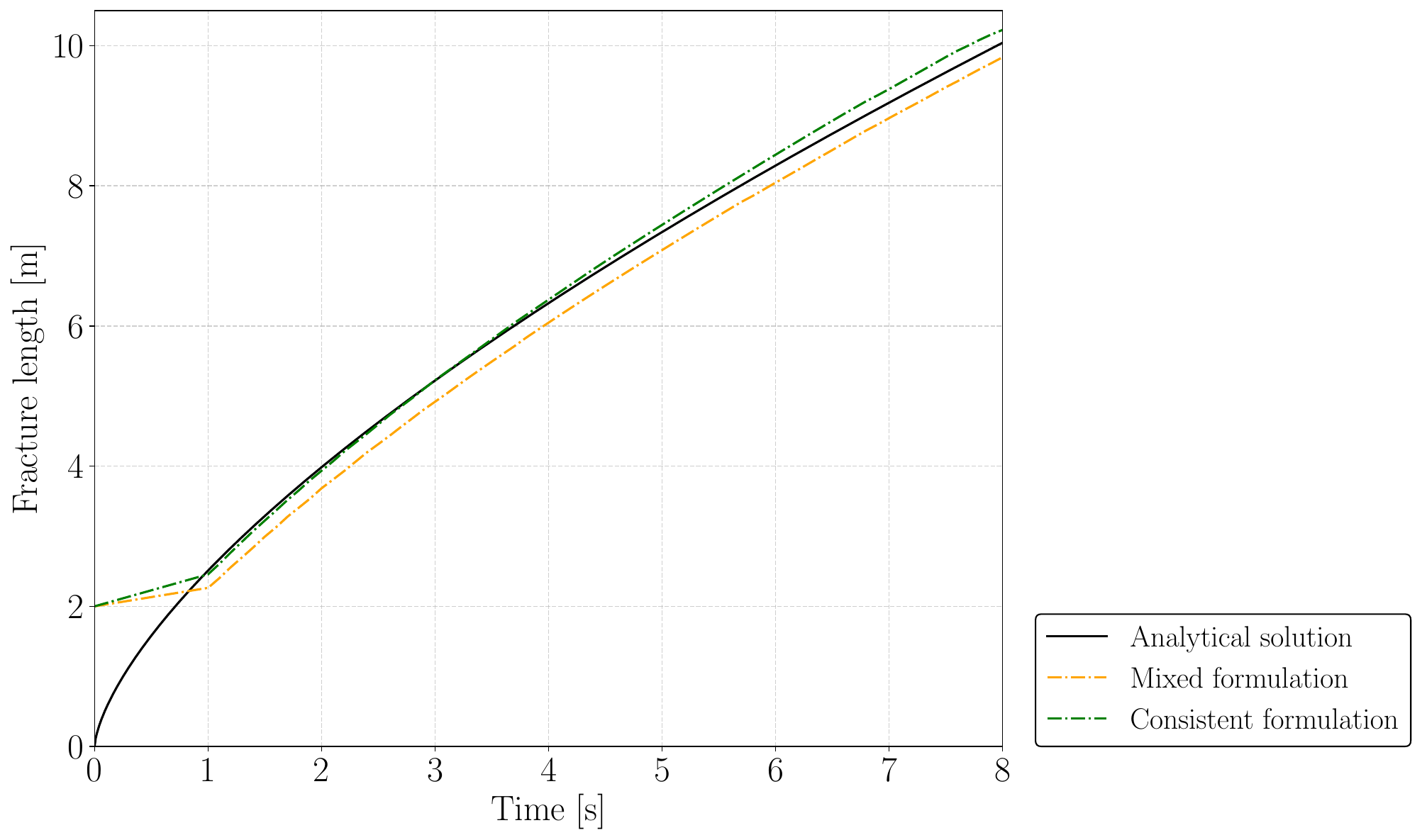}
        \caption{}
        \label{fig:KGD-length}
    \end{subfigure}
    \caption{Comparisons of (a) pressure at the injection point, (b) fracture aperture at the injection point, and (c) fracture length for the KGD hydraulic fracturing. The black solid lines are the analytical solutions, while orange and green dashed lines represent simulation results from the consistent and mixed formulations.}
    \label{fig:KGD_results}
\end{figure}

Fig.~\ref{fig:KGD_results} compares the analytical solutions with the numerical results obtained using the proposed consistent model and the mixed formulation from~\citep{yi2020consistent, you_poroelastic_2023}. 
Compared with the mixed formulation, our consistent formulation yields a lower fracture initiation pressure at the injection point (Fig.~\ref{fig:KGD_results}a), showing closer alignment with the analytical solution. 
This indicates that the pressure-deformation coupling term in the consistent formulation effectively prevents an overestimation of the fluid pressure during fracturing. 
Improvement over the mixed formulation is also evident in the comparative evaluation of the fracture aperture at the injection point and the fracture length (Figs.~\ref{fig:KGD_results}b and~\ref{fig:KGD_results}c). 

\subsection{Fluid driven fracturing in poro-elastoplastic model}
\label{sec:HF in poro elastoplastic}

\begin{figure}[h!]
    \centering
    \begin{subfigure}{0.49\textwidth}
        \centering
        \includegraphics[width=\textwidth]{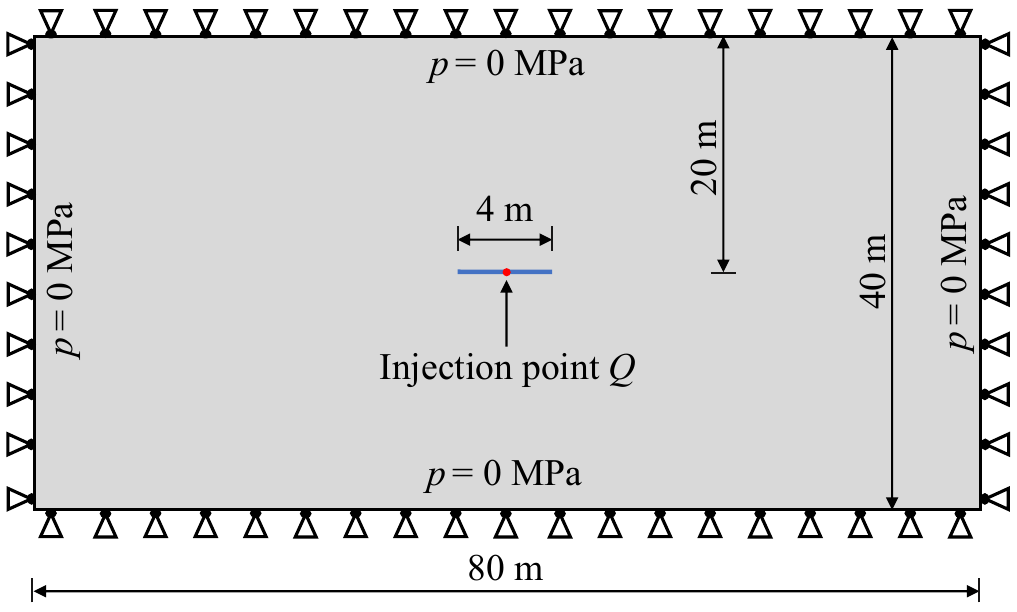}
        \caption{}
    \end{subfigure}
    \hfill 
    \begin{subfigure}{0.49\textwidth}
        \centering
        \includegraphics[width=\textwidth]{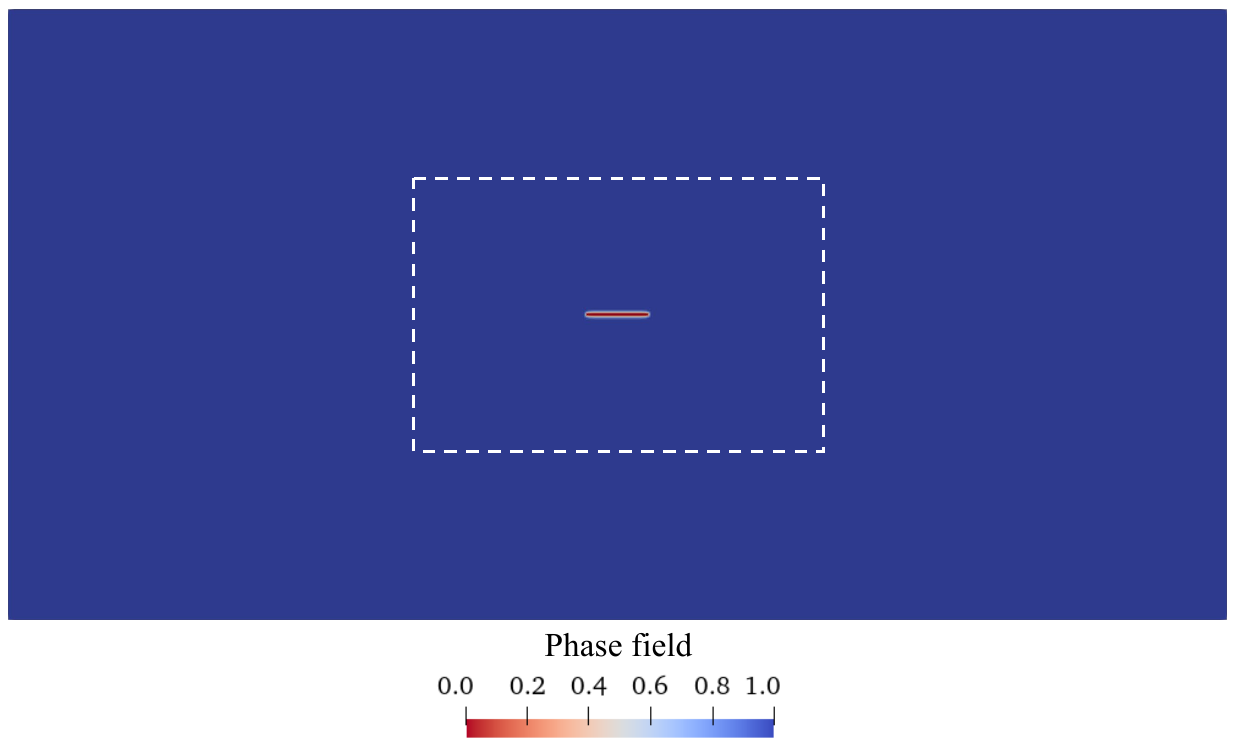}
        \caption{}
    \end{subfigure}
    \caption{A schematic of the geometry and boundary conditions of the hydraulic fracturing model (a). A close-up of the region bounded by the white dashed box is presented in the following figure to highlight the evolution of multiple fields (b).}
    \label{fig:injection-model}
\end{figure}

Following the verification with the elastic formulation in Sec.~\ref{sec:verification_KGD}, this section compares simulation results obtained using the poro-elastic and poro-elastoplastic models in a 2D domain ($[0~\mrm{m},\, 80~\mrm{m}] \times [0~\mrm{m},\, 40~\mrm{m}]$) with an initial crack located at $[38~\mrm{m},\, 42~\mrm{m}] \times \{20~\mrm{m}\}$ (Fig.~\ref{fig:injection-model}). 
The four edges are permeable with a constant pressure of 0~MPa, and the displacements are fixed in their normal directions. 
A locally refined mesh of 31,270 quadrilateral elements was employed for the discretization, with a minimum element size of $h = 0.1$ m. 
The Young's modulus and Poisson's ratio of the material are the same as those in Table~\ref{tab:hf properties}, while $G_c = 100$ N/m, $\alpha_0 = 0.6$, $\phi_0 = 0.01$, and $K_\mathrm{m} = 1 \times 10^{-15} \, \mathrm{m}^2$. 
For this example, fluid injection with a higher viscosity ($\mu = 1 \times 10^{-5}$ Pa$\cdot$s) was simulated at an injection rate of $Q_\mathrm{inj} = 5 \times 10^{-4} \, \mathrm{m}^2 / \mathrm{s}$ with a time step size of $\Delta t = 0.2$ s. 
The remaing parameters are the same as those in Table~\ref{tab: parameter_comparison} except for $A = 0.905$.

\begin{figure}[h!]
    \centering
    \includegraphics[width=1\linewidth]{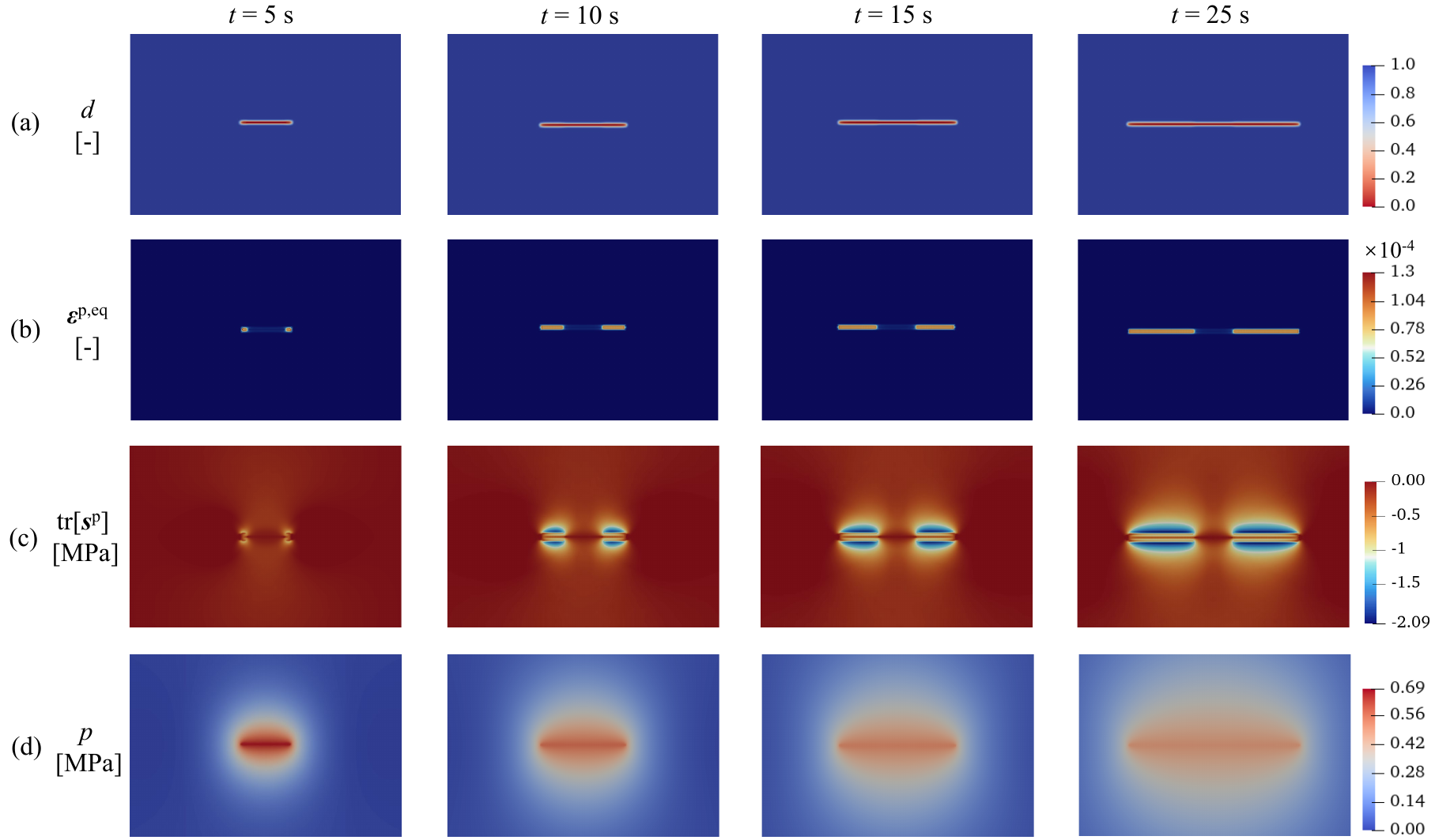}
    \caption{Simulation results of hydraulic fracturing in the elastoplastic model at $t$ = 5, 10, 15, and 25~s: (a) phase-field $d$, (b) equivalent plastic strain $\bm{\varepsilon}^\mathrm{p, eq}$, (c) trace of local stress $\mathrm{tr}[\bm{s}^\mathrm{p}]$, and (d) pressure $p$.}
    \label{fig:plastic injection}
\end{figure}

Fig.~\ref{fig:plastic injection} presents the evolution of the phase field, plastic strain, the trace of the local stress tensor, and pore pressure. 
The mean local stress remains compressive ($\mathrm{tr}[\bm{s}^\mathrm{p}] < 0$) in the region adjacent to the propagating fracture (Fig.~\ref{fig:plastic injection}c), and plastic deformation extends over a finite width on both sides of the propagating fracture, forming a sheath-like configuration around it (Fig.~\ref{fig:plastic injection}b). 
At the crack tip, however, the local stress is tensile, confirming that the fracture propagates in a tension-dominated mode driven by the fluid pressure (Fig.~\ref{fig:plastic injection}d).

\begin{figure}[h!]
    \centering
    \begin{subfigure}{0.49\textwidth}
        \centering
        \includegraphics[width=\textwidth]{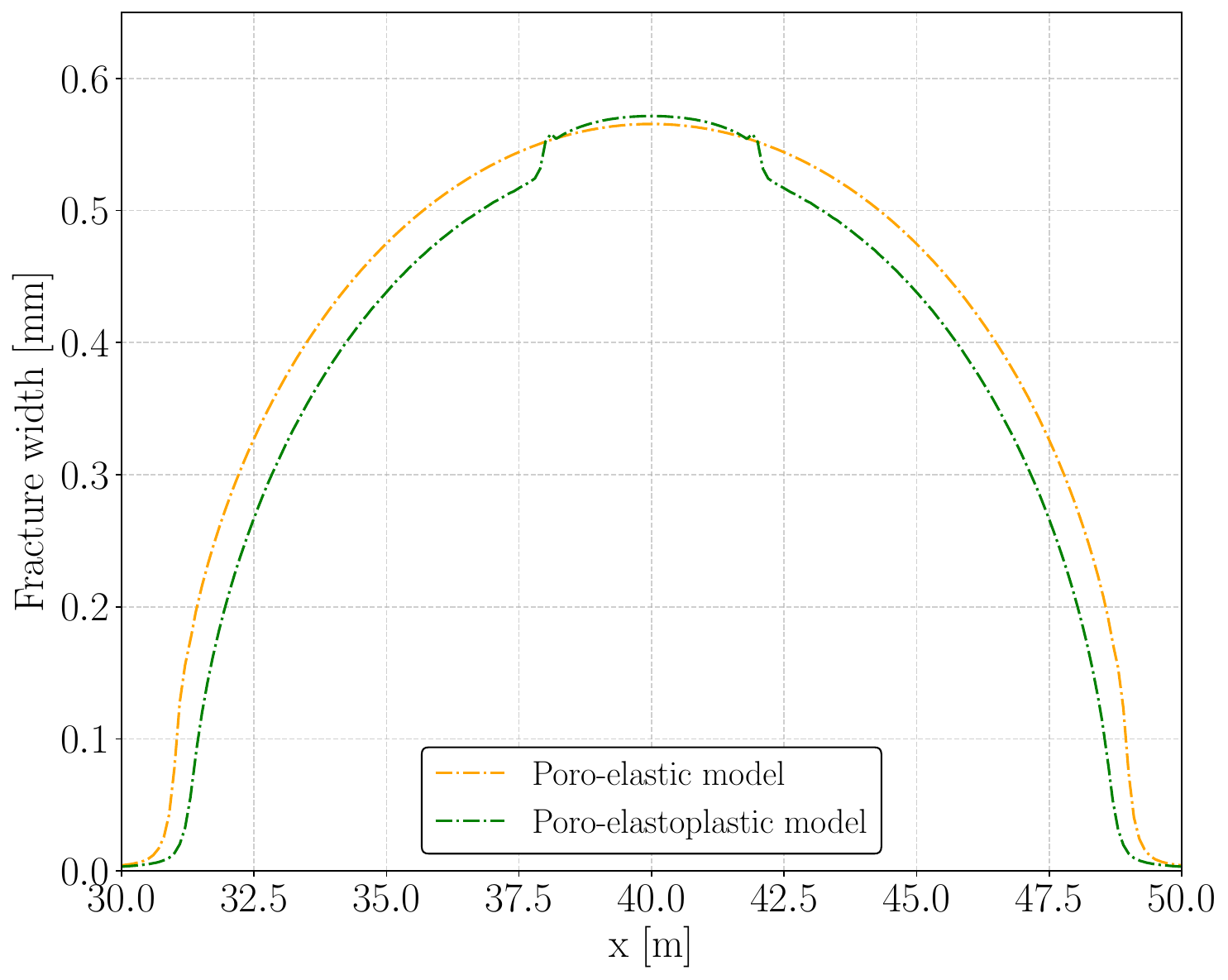}
        \caption{}
    \end{subfigure}
    \hfill 
    \begin{subfigure}{0.49\textwidth}
        \centering
        \includegraphics[width=\textwidth]{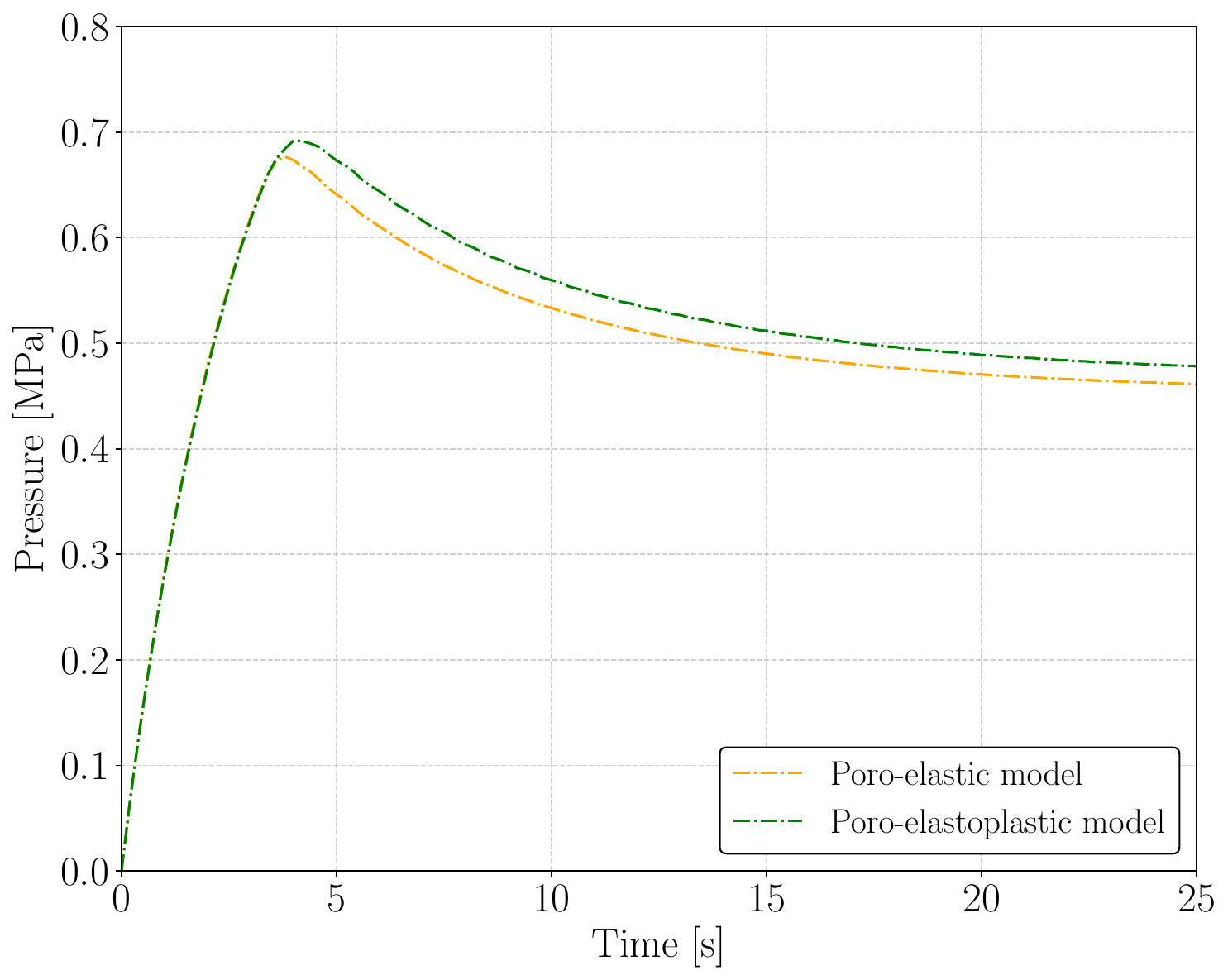}
        \caption{}
    \end{subfigure}
    \caption{Comparisons of (a) fracture width at $t= 25.0$~s and (b) pressure evolution at the injection point between the poro-elastic and the poro-elastoplastic models.}
    \label{fig:constitutive-comparison}
\end{figure}

Comparing the fracture width profiles between the elastic and elastoplastic cases, Fig.~\ref{fig:constitutive-comparison}a shows that plastic deformation restricts fracture opening, yielding a smaller width than in the elastic case. 
Similar fracture width profiles have been reported in existing literature employing poro-elastoplastic models~\citep{papanastasiou1997influence, kienle2022variational}. 
The evolution of injection pressure also highlights the additional plastic dissipation during fracture propagation (Fig.~\ref{fig:constitutive-comparison}b). 
At the onset of fracture propagation, plastic deformation near the crack tip provides extra resistance, leading to a higher breakdown pressure. 
Once propagation begins, a higher pore pressure is required within the crack to sustain this additional plastic dissipation. 
Evidently, the fracturing process is accompanied by significant plastic energy dissipation, resulting in hydromechanical coupling behavior that differs substantially from predictions based on a purely poroelastic formulation.

\subsection{Compressive-shearing fracture in quasi-brittle porous media}
\label{sec:cs}

\begin{figure}[h!]
    \centering
    \includegraphics[width=0.3\linewidth]{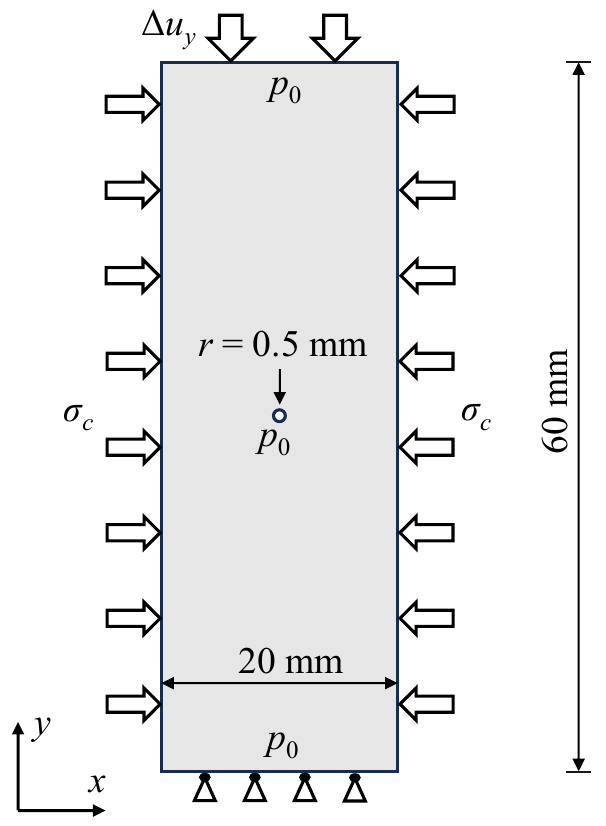}
    \caption{A schematic representation of the biaxial compression model}
    \label{fig:bc model}
\end{figure}

Lastly, we present a biaxial compression test on a perforated 2D domain under plane strain conditions (Fig.~\ref{fig:bc model}). 
A uniform mesh of 35,460 quadrilateral elements was employed with a minimum element size $h = 0.2$~mm.
The left and right boundaries were set to impermeable, while permeable conditions were prescribed at the top, bottom, and internal hole, where the pressure was maintained at the initial pressure $p_0$ (Table~\ref{tab: parameters in biaxial compression}). 
The bottom boundary of the model was fixed in the normal direction. 
Loading was applied with two stages: in the first stage, the confining stress was applied by uniformly increasing the stress on the lateral and top boundaries until reaching $\sigma_c = 5$ MPa. 
In the second stage, the confining pressure was kept constant while a displacement of 0.005 mm per time step was imposed on the top boundary. 
The simulation parameters are listed in Table~\ref{tab: parameters in biaxial compression}, and the plasticity parameters are the same as those in Sec~\ref{sec:HF in poro elastoplastic}.  

\begin{table}[h!]
    \centering
    \caption{Parameters for the biaxial compression simulation}
    \label{tab: parameters in biaxial compression}
    \begin{tabular}{p{5cm}cc}
    \toprule
        Properties & Value & Unit\\
        \midrule
        Young's modulus ($E$) & 49 & GPa \\
        Poisson's ratio ($\upsilon$) & 0.19 & -\\
        Critical energy release rate ($G_c$) & 300 & N/m\\
        Biot's coefficient ($\alpha_0$) & 0.9 & -\\
        Porosity ($\phi_0$) & 0.01 & - \\
        Fluid compressibility ($c_f$) & 0.0 & - \\
        Permeability ($K_\mathrm{m}$) & $5 \times 10^{-14}$ & $\text{m}^2$\\
        Fluid viscosity ($\mu$) & $1 \times 10^{-3}$ & Pa s \\
        Friction coefficient & 0.905 & - \\
        Initial pressure ($p_0$) & 0, 1, 2 & MPa \\
        \bottomrule
    \end{tabular}\\
\end{table}

\begin{figure}[h!]
    \centering
    \includegraphics[width=0.96\linewidth]{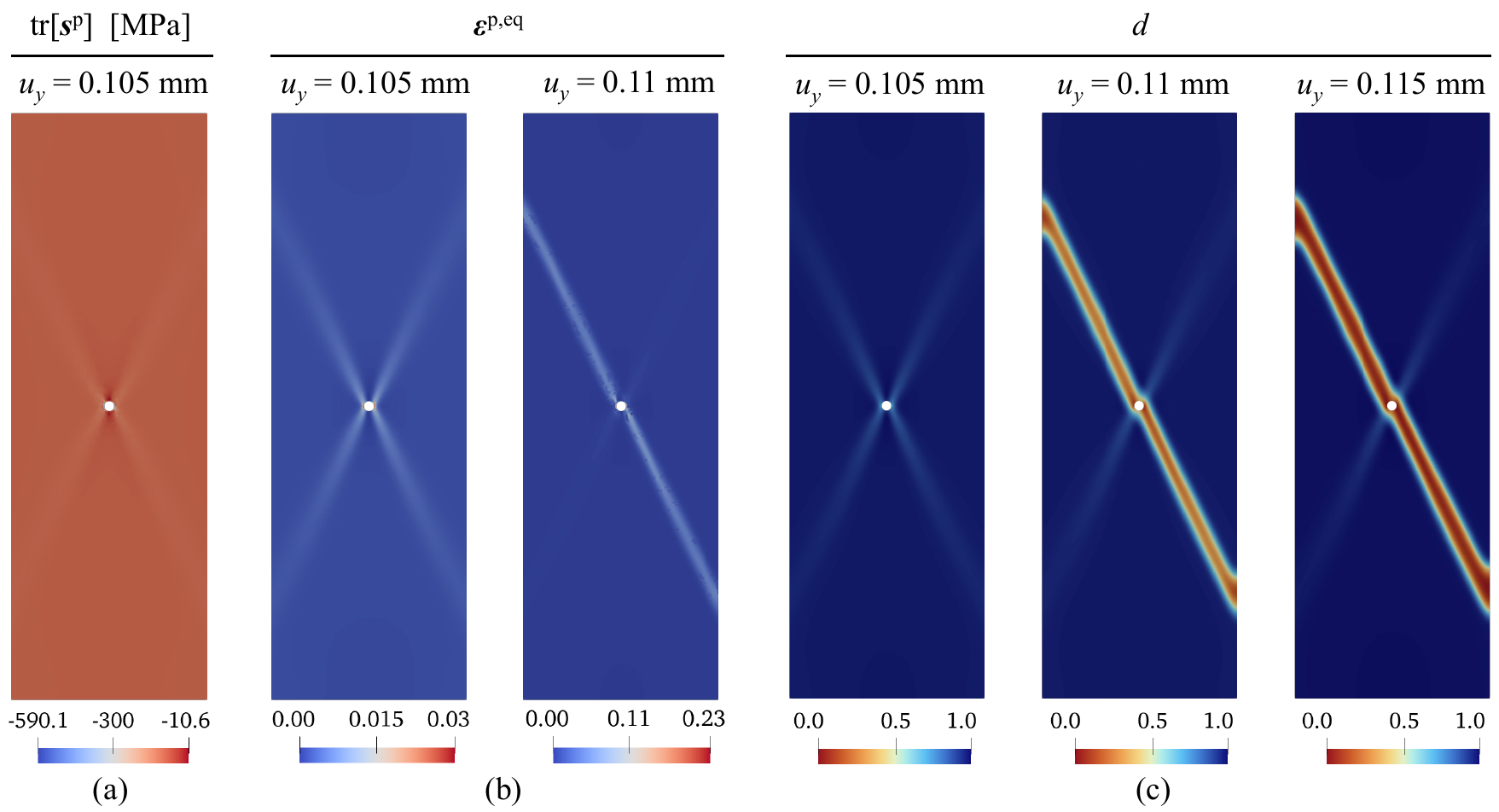}
    \caption{Fracture process of the perforated sample with initial pressure $p_0 = 1$ MPa, where (a) trace of local stress $\mathrm{tr}[\bm{s}^\mathrm{p}]$ pre-fracture, (b) equivalent plastic strain ${\varepsilon}^\mathrm{p,eq}$ pre- and post-fracture, and (c) phase-field pre- and post-fracture are presented.}
    \label{fig:fracture process}
\end{figure}

The simulation results with initial pressure $p_0 = 1$ MPa are shown in Fig.~\ref{fig:fracture process}. 
At $u_y=0.105$~mm, the entire sample is subjected to the compressive-shear regime ($\mathrm{tr}[\bm{s}^\mathrm{p}] < 0$) (Fig.~\ref{fig:fracture process}a). 
At this loading stage, the equivalent plastic strain ${\varepsilon}^\mathrm{p,eq}$ initiates at the perforation and localizes along an inclined band centered at the hole, forming a shear band (Fig.~\ref{fig:fracture process}b). 
With evolving plastic strain and damage, a fracture nucleates at the hole (Fig.~\ref{fig:fracture process}c). 
As the load further increases, a typical shear (mode II) fracture forms with a narrow shear band that overlaps with the damaged zone.

\begin{figure}[h!]
    \centering
    \includegraphics[width=0.96\linewidth]{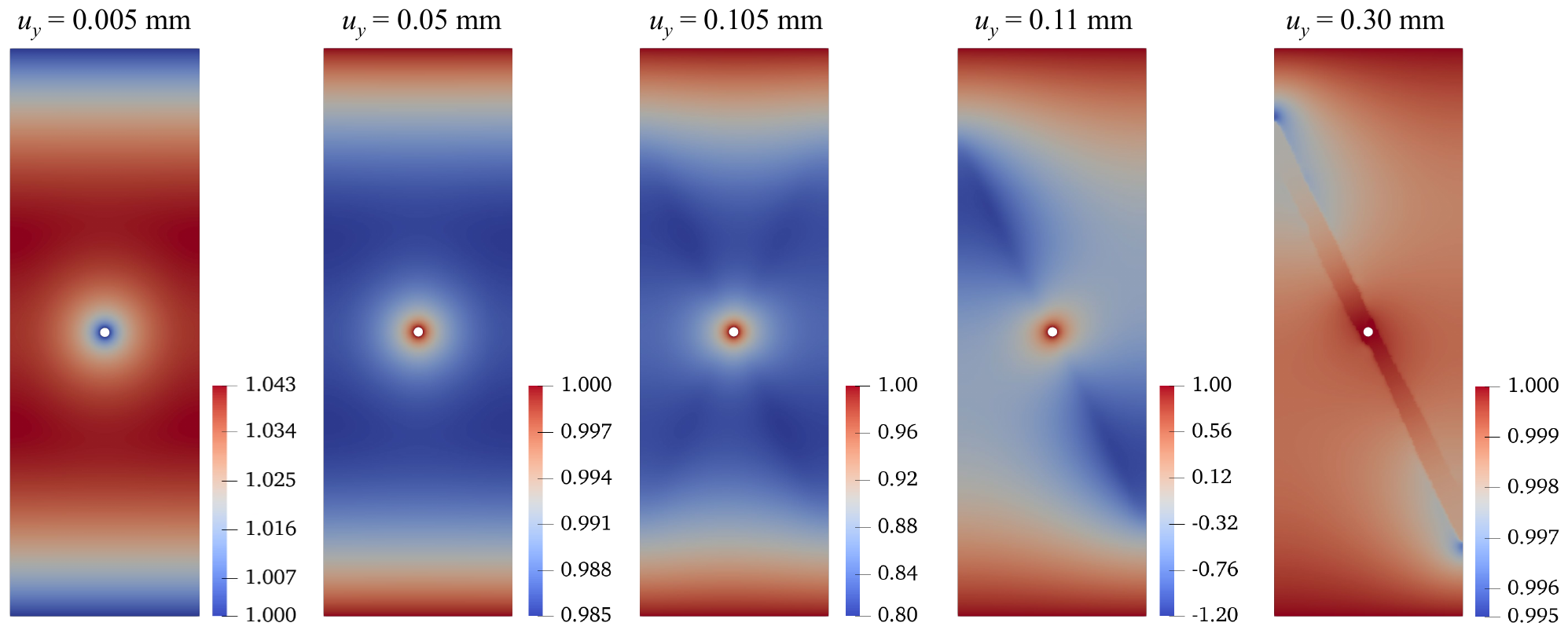}
    \caption{Pressure evolution during biaxial compression. (Unit: MPa)}
    \label{fig:fracture pressure}
\end{figure}

Fig.~\ref{fig:fracture pressure} shows the pressure profiles during the evolution of damage and plasticity. 
At the initial stage of loading ($u_y = 0.005$ mm), the pore pressure in the sample increases due to compression. 
As the load steadily increases ($u_y = 0.05$ mm), this overpressure gradually dissipates. 
Prior to fracture nucleation ($u_y = 0.105$ mm), the increased porosity (volumetric dilation) induced by plastic frictional deformation causes the pore pressure to drop below its initial value.
This localized pressure reduction draws fluid from the surrounding porous matrix into the damaged zone. 
In the post-fracture stage ($u_y = 0.11$ mm), the permeability within the shear band increases dramatically, forming a high-conductivity flow path along the fracture where negative pore pressure develops, accelerating fluid discharge from the matrix. 
With further loading ($u_y = 0.30$ mm), the pore pressure within the matrix eventually returns to the initial level ($p = 1$ MPa), reaching a steady state. 
A similar evolution of the pressure profile has also been reported in~\citet{callari2002finite,hadzalic2018failure, ULLOA2022115084}.

\begin{figure}[h!]
    \centering
    \begin{subfigure}{0.49\textwidth}
        \centering
        \includegraphics[width=\textwidth]{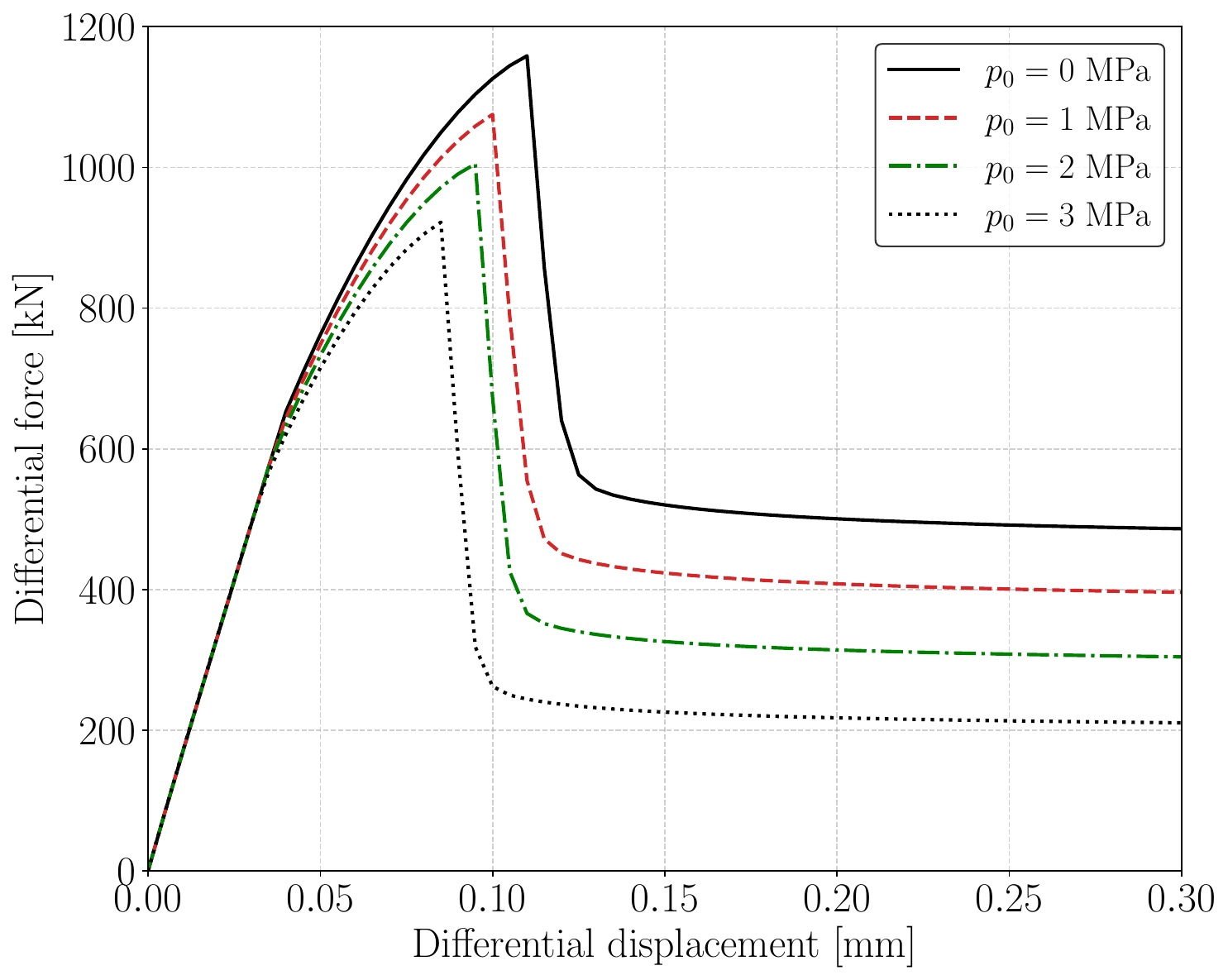}
        \caption{}
    \end{subfigure}
    \hfill 
    \begin{subfigure}{0.49\textwidth}
        \centering
        \includegraphics[width=\textwidth]{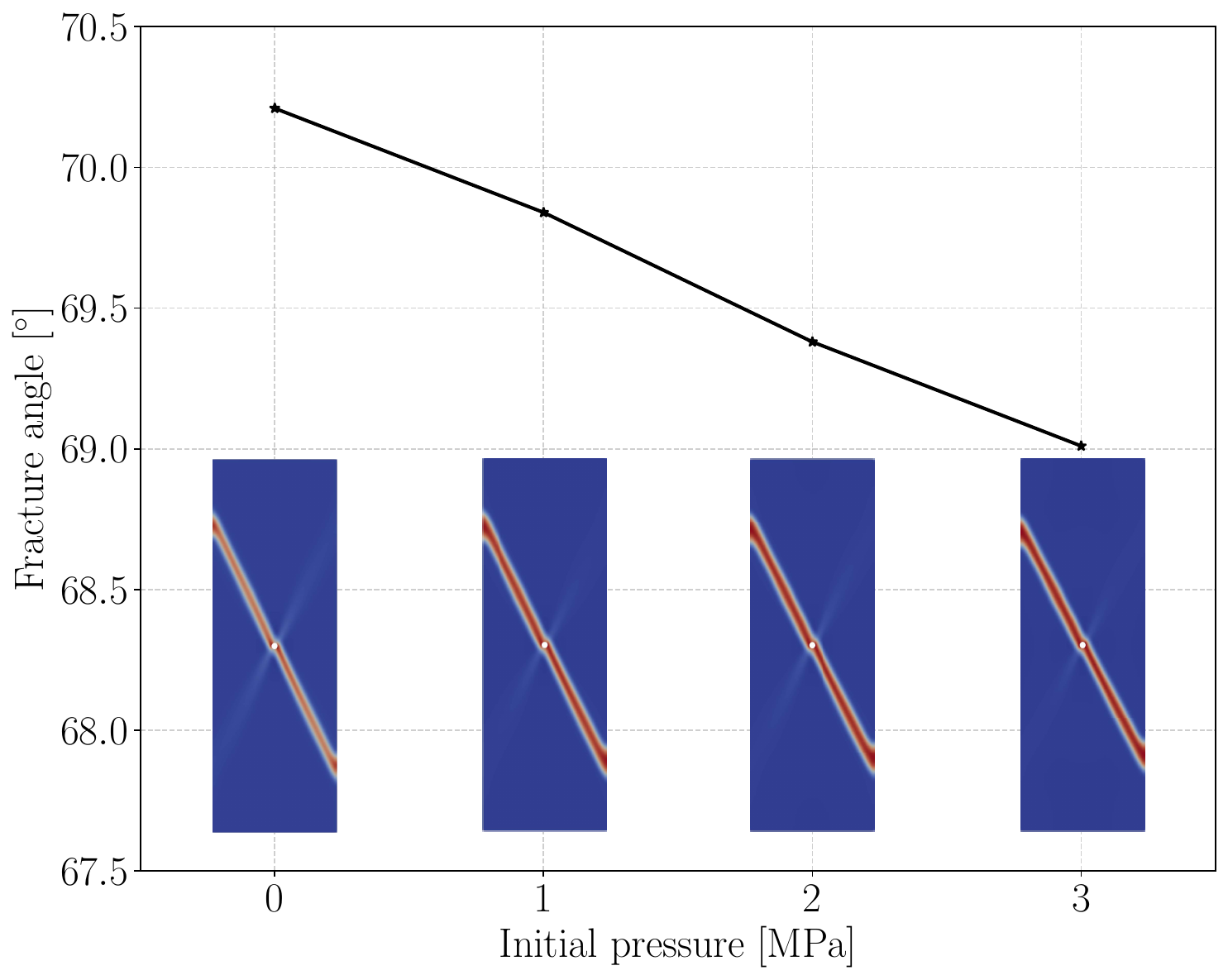}
        \caption{}
    \end{subfigure}
    \caption{Comparison of the post-fracture features: (a) Differential displacement vs. differential force curve, and (b) fracture angles. We consider a unit thickness (1 m) under the assumption of plane strain.}
    \label{fig:disp vs f}
\end{figure}

Fig.~\ref{fig:disp vs f}a shows the differential force-displacement curves at the top boundary under various initial pressures ($p_0$=0, 1, 2, and 3~MPa). 
In all cases, the force-displacement curve increases linearly during the early stage of loading, indicating elastic deformation. 
Subsequently, due to the accumulation of damage and plastic deformation, the curve exhibits a hardening segment until the peak loading is reached. 
Samples with lower initial pressure $p_0$ sustain higher peak loads, which is consistent with experimental results in~\citet{ZHU2023103789}. 
At the post-peak stage, the residual strength of the sample $F_\mathrm{res}$ is determined by the friction coefficient and confining stresses. 
The pore pressure weakens this residual strength and shows an approximately linear relation between the pore pressure increment $\Delta p_0$ and the decrease in residual strength $\Delta F_\mathrm{res}$. 
Fig.~\ref{fig:disp vs f}b shows the decrease in fracture angle with higher initial pore pressure. 
This variation, though slight, demonstrates that the pore pressure can promote the transition from a pure shear fracture to the hybrid fracture.  

\section{Conclusion}
\label{Conclusion}

This paper revisits the hydromechanical formulation of the micromechanics-based phase-field model in poro-elastoplastic media and presents the consistent expression for the phase-field driving force. 
Furthermore, our analyses show that the analytically derived strength surface is continuous across the tension-compression transition when employing an associative plastic flow rule. 
The proposed micromechanics-based hydromechanical framework is fully compatible with the previously proposed stress-dependent cohesive degradation function~\citep{li2025cohesive} and provides excellent agreement with experimental results for hydromechanically coupled fracture. 
The numerical results for the KGD benchmark in a poroelastic material demonstrate that our consistent formulation improves the accuracy of fluid-driven fracturing simulations. 
Finally, hydraulic fracturing and biaxial compression tests in poro-elastoplastic media demonstrate that the proposed model is capable of capturing both mechanically induced shear fractures and hydraulically induced tensile fractures, showing strong potential for applications in geoenergy production and subsurface engineering.

	
	
\section*{Acknowledgment}
The authors from Tongji University first acknowledge the funding support by the National Natural Science Foundation of China Original Exploration Program under Grant No. 4255000075. This research was funded in whole or in part by the Austrian Science Fund (FWF)  10.55776/PIN9246524.

\appendix


\section{Consistent hardening modulus}
\label{Consistent hardening modulus}
This Appendix shows the derivation of the degradation function of the hardening modulus (Eq.~\eqref{eq: gp}), using the continuities of the stresses ($\bm{\sigma}_\mrm{open} = \bm{\sigma}_\mrm{close}$) and the local stresses ($\bm{s}_{\mathrm{open}}=\bm{s}_{\mathrm{close}}$) at the opening-closure transition.
To ensure the continuity, Eqs.\eqref{stress-strain relation} and \eqref{generalized stress inelastic} impose the following equalities:
\begin{align}
    \label{eq: stress_continuity}
    \mathbb{C}_\mathrm{dam}:\bm{\varepsilon} - \alpha p \mathbf{I} - \mathbb{C}:\left(\bm{\varepsilon} - \bm{\varepsilon}^\mathrm{p} \right) + \alpha_0 p \mathbf{I} = 0  \\
    \label{eq: local_stress_continuiity}
    \mathbb{C}:\left(\bm{\varepsilon} - \bm{\varepsilon}^\mathrm{p} \right) - \mathbb{H}:\bm{\varepsilon}^\mathrm{p} + (1 - \alpha_0)p\mathbf{I} = 0
    .
\end{align}
Solving Eq.~\ref{eq: local_stress_continuiity} for $\bm{\eps}$ yields
\begin{equation}
\label{plastic strain HM}
    \bm{\varepsilon} = \mathbb{C}^{-1}:(\mathbb{H} + \mathbb{C}):\bm{\varepsilon}^\mathrm{p} - \mathbb{C}^{-1}:(1 - \alpha_0)p\mathbf{I}
    .
\end{equation}
Substituting Eq.\eqref{plastic strain HM} into Eq.~\eqref{eq: stress_continuity}, we have
\begin{equation}
    \begin{aligned}
    \label{eq: stress_continuity_2}
        \left[g(d) - 1 \right] \left[(\mathbb{H} + \mathbb{C}):\bm{\varepsilon}^\mathrm{p} - (1 - \alpha_0)p\mathbf{I} \right] - \alpha p \mathbf{I} + \mathbb{C}:\bm{\varepsilon}^\mathrm{p} + \alpha_0 p\mathbf{I} = 0
        .
    \end{aligned}
\end{equation}
Rearranging Eq.~\eqref{eq: stress_continuity_2}, we obtain
\begin{equation}
    \begin{aligned}
        \left[ (g(d) - 1) \mathbb{H} + g(d) \mathbb{C} \right]:\epsp - [(g(d) - 1)(1 - \alpha_0) - \alpha + \alpha_0]p\mathbf{I} = 0
        .
    \end{aligned}
\end{equation}
Note that the above equation should be satisfied for arbitrary $\epsp$ and $p$. 
The second term is 0 because of Eq.~\eqref{eq:biot}-1.
Then for the first term, the following should hold
\begin{align}
    (g(d) - 1) \mathbb{H} + g(d) \mathbb{C} = 0
    .
\end{align}
Thus we have
\begin{align}
     \mathbb{H} = \frac{g(d)}{1 - g(d)}\mathbb{C}
    \label{eq:strong form-1} 
\end{align}

\section{Temporal and spatial discretizations of the governing equations}
\label{Temporal and spatial discretizations of the governing equations}

We apply the back Euler scheme at step $n$ for the temporal discretization: 
\begin{equation}
\label{eq: back euler}
    \begin{aligned}
        \frac{\partial \mathrm{tr}[\bm{\varepsilon}]}{\partial t} = \frac{\mathrm{tr}[\bm{\varepsilon_n - \bm{\varepsilon}_{n-1}}]}{\Delta t}, \, \frac{\partial \mathrm{tr}[\epsp]}{\partial t} = \frac{\mathrm{tr}[\epsp_n - \epsp_{n-1}]}{\Delta t}, \, \frac{\partial p}{\partial t} = \frac{p_n - p_{n-1}}{\Delta t}, \, \frac{\partial d}{\partial t} = \frac{d_n - d_{n-1}}{\Delta t}
    \end{aligned}
\end{equation}
Substituting Eq.\eqref{eq: back euler} into Eq.~\eqref{eq: govern fluid} gives
\begin{equation}
\label{eq: time derivate of fluid content}
    \begin{aligned}
         \frac{\partial \xi(\bm{\varepsilon}, \epsp)}{\partial t} =
    \begin{dcases}
        \alpha \frac{\mathrm{tr}[\bm{\varepsilon}_n - \bm{\varepsilon}_{n-1}]}{\Delta t} + \frac{p_n - p_{n-1}}{M\Delta t} \quad &\text{if} \quad \mathrm{tr}[\bm{s}^\mathrm{p}]=0 \\
        \alpha_0 \frac{\mathrm{tr}[\bm{\varepsilon}_n - \bm{\varepsilon}_{n-1}]}{\Delta t}  + \frac{p_n - p_{n-1}}{M_0 \Delta t} 
         - (\alpha_0 - 1)\frac{\mathrm{tr}[\bm{\varepsilon}^\mathrm{p}_n - \bm{\varepsilon}^\mathrm{p}_{n-1}]}{\Delta t} \quad &\text{if} \quad \mathrm{tr}[\bm{s}^\mathrm{p}]<0
    \end{dcases}
    \end{aligned}
\end{equation}

The primary variables are spatially discretized with shape functions $\bm{N}_i$ and $N_i$, and transformation matrix $\bm{B}^u_i$ and $\bm{B}_i$:
\begin{equation}
    \begin{aligned}
        &\bm{u} = \bm{N}_i \bm{u}_i, \quad d = N_i d_i, \quad p = N_i p_i  \\
        &\nabla\bm{u} = \bm{B}^u_i \bm{u}_i, \quad \nabla d = \bm{B}_i d_i, \quad \nabla p = \bm{B}_i p_i
    \end{aligned}
\end{equation}
where $i$ denotes the nodes in an element, e.g., $i=1,2,3,4$ in a first-order quadrilateral element. 
Then the residuals take the form

\begin{equation}
\label{eq: rhs_p}
     r_i^p =
    \begin{dcases}
        \int_{\Omega} (N_i)^\mathrm{T} \left(  \alpha \frac{\mathrm{tr}[\bm{\varepsilon}_n - \bm{\varepsilon}_{n-1}]}{\Delta t} + \frac{p_n - p_{n-1}}{M\Delta t} \right) \mathrm{d}V \\
        \qquad + \int_\Omega (\bm{B}_i)^\mathrm{T} \frac{\bm{K}}{\nu}\nabla p \mathrm{d}V - \int_\Omega (N_i)^\mathrm{T}  Q \mathrm{d}V + \int_{\partial \Omega_q} (N_i)^\mathrm{T} \bar{\bm{q}} \mathrm{d}S  \quad \text{if} \quad \mathrm{tr}[\bm{s}^\mathrm{p}] = 0  \\
        \int_{\Omega} (N_i)^\mathrm{T} \left( \alpha_0 \frac{\mathrm{tr}[\bm{\varepsilon}_n - \bm{\varepsilon}_{n-1}]}{\Delta t}  + \frac{p_n - p_{n-1}}{M_0 \Delta t}- (\alpha_0 - 1)\frac{\mathrm{tr}[\bm{\varepsilon}^\mathrm{p}_n - \bm{\varepsilon}^\mathrm{p}_{n-1}]}{\Delta t} \right) \mathrm{d}V \\
        \qquad + \int_\Omega (\bm{B}_i)^\mathrm{T} \frac{\bm{K}}{\nu}\nabla p \mathrm{d}V - \int_\Omega (N_i)^\mathrm{T}  Q \mathrm{d}V + \int_{\partial \Omega_q} (N_i)^\mathrm{T} \bar{\bm{q}} \mathrm{d}S  \quad \text{if} \quad \mathrm{tr}[\bm{s}^\mathrm{p}] < 0 
    \end{dcases}
\end{equation}
\begin{equation}
\label{eq:rhs-u}
    \bm{r}_i^u =
    \begin{dcases}
        \int_{\Omega} (\bm{B}^u_i)^\mathrm{T} \left( \mathbb{C}_\mathrm{dam}(d):\bm{\varepsilon} - \alpha p \mathbf{I} \right) \mathrm{d}V - \int_{\Omega}(\bm{N}_i)^\mathrm{T} \bm{b} \mathrm{d}V -  \int_{\partial\Omega_t}(\bm{N}_i)^\mathrm{T} \bar{\bm{t}}\mathrm{d}S \quad &\text{if} \quad \mathrm{tr}[\bm{s}^\mathrm{p}] = 0\\
        \int_{\Omega}(\bm{B}^u_i)^\mathrm{T} \left( \mathbb{C}:(\bm{\varepsilon} - \bm{\varepsilon}^\mathrm{p}) - \alpha_0 p \mathbf{I} \right) \mathrm{d}V - \int_{\Omega}(\bm{N}_i)^\mathrm{T} \bm{b} \mathrm{d}V -  \int_{\partial\Omega_t}(\bm{N}_i)^\mathrm{T}  \bar{\bm{t}}\mathrm{d}S \quad &\text{if} \quad \mathrm{tr}[\bm{s}^\mathrm{p}] < 0
    \end{dcases}
\end{equation}
\begin{equation}
\label{eq:rhs-d}
    r_i^d = 
    \begin{dcases}
        \int_{\Omega} (N_i)^\mathrm{T} \dfrac{1}{2}\dfrac{\partial g(d)}{\partial d}\left[ \bm{\varepsilon}:\mathbb{C}:\bm{\varepsilon} + \dfrac{(1 - \alpha_0)^2}{K}p^2 + 2(1 - \alpha_0)p\mathrm{tr}\bm{\varepsilon} \right] \mathrm{d}V \\
        \qquad - \int_{\Omega} (N_i)^\mathrm{T} \dfrac{2G_c d}{\pi \ell} \mathrm{d}V + \int_\Omega \dfrac{2G_c \ell}{\pi}(\bm{B}_i)^\mathrm{T} \nabla d \,\mathrm{d}V  \quad \text{if} \quad \mathrm{tr}[\bm{s}^\mathrm{p}] = 0  \\
        \int_{\Omega} (N_i)^\mathrm{T} \dfrac{1}{2}\dfrac{\partial g_p(d)}{\partial d}\epsp:\mathbb{C}:\epsp \, \mathrm{d}V \\
        \qquad - \int_{\Omega} (N_i)^\mathrm{T} \dfrac{2G_c d}{\pi \ell} \mathrm{d}V + \int_\Omega \dfrac{2G_c \ell}{\pi}(\bm{B}_i)^\mathrm{T} \nabla d \, \mathrm{d}V  \quad \text{if} \quad \mathrm{tr}[\bm{s}^\mathrm{p}] < 0 
    \end{dcases}
\end{equation}

When $\mathrm{tr}[\bm{s}^\mathrm{p}] < 0$, the plastic strain $\bm{\varepsilon}^\mathrm{p}$ becomes dependent on the primary variables $p$, and $\bm{u}$. Consequently, terms involving plastic strain need to be differentiated with respect to these primary variables to obtain the complete Jacobian matrices, which give
\begin{equation}
    \begin{aligned}
        \frac{\partial \bm{\varepsilon}^\mathrm{p}}{\partial p} =& \frac{\partial (\Delta \lambda \mathbf{D})}{\partial p} = \frac{\partial \Delta \lambda }{\partial p}\mathbf{D} + \Delta \lambda\frac{\partial \mathbf{D}}{\partial p} = \zeta \frac{A \mathrm{tr}\mathbf{I}}{3}(1 - \alpha_0)\mathbf{D} ,\\
        \frac{\partial \bm{\varepsilon}^\mathrm{p}}{\partial \bm{\varepsilon}} =& \zeta \mathbf{D} \otimes (2\mu \mathbf{V} + KA \mathbf{I}) + \frac{2\mu \Delta \lambda}{\| \bm{s}^\mathrm{p \, trial}_\mathrm{dev} \|}(\mathbb{I} - \frac{1}{3}\mathbf{I}\otimes\mathbf{I} - \mathbf{V}\otimes\mathbf{V}).
    \end{aligned}
\end{equation}

Accordingly, the Jacobian matrices can be constructed by taking the partial derivatives of the residuals with respect to the associated primary variables. 
Note that as we employ a monolithic scheme for the coupled hydromechanical process, the Jacobian $\mathrm{K}^{pu}$ and $\mathrm{K}^{up}$ should also be provided.

\begin{equation}
\label{eq: jacobian-pu}
    \begin{aligned}
        \mathbf{K}^{pp}_{ij} &= \frac{\partial r_i^p}{\partial p_j} = 
        \begin{dcases}
         \int_{\Omega} \frac{1}{M \Delta t} (N_i)^\mathrm{T} N_i  \mathrm{d}V + \int_{\Omega} (\bm{B}_i)^\mathrm{T} \frac{\mathbf{K}}{\nu}\bm{B}_i \mathrm{d}V \quad &\text{if} \quad \mathrm{tr}[\bm{s}^\mathrm{p}] = 0 \\
        \int_{\Omega} \frac{1}{M_0 \Delta t} (N_i)^\mathrm{T} N_i \mathrm{d}V + \int_{\Omega} (\bm{B}_i)^\mathrm{T} \frac{\mathbf{K}}{\nu}\bm{B}_i \mathrm{d}V \\  
        \qquad + \int_\Omega  \frac{(1 - \alpha_0)^2}{\Delta t} \left( \frac{A \mathrm{tr}\mathbf{I}}{3}\right)^2 \zeta (N_i)^\mathrm{T} N_i\mathrm{d}V \quad &\text{if} \quad \mathrm{tr}[\bm{s}^\mathrm{p}] < 0 
        \end{dcases}
    \end{aligned}
\end{equation}
\begin{equation}
    \begin{aligned}
        \mathbf{K}^{pu}_{ij} &= \frac{\partial r_i^p}{\partial \bm{u}_j} = 
        \begin{dcases}
        \int_{\Omega} (N_i)^\mathrm{T} \left( \frac{\alpha}{\Delta t} \mathbf{I} \bm{B}_i^u  \right) \mathrm{d}V &\quad \text{if} \quad \mathrm{tr}[\bm{s}^\mathrm{p}] = 0  \\
        \int_{\Omega} (N_i)^\mathrm{T} \left( \frac{\alpha_0}{\Delta t} \mathbf{I} \bm{B}_i^u \right) \mathrm{d}V \\
        \qquad + \int_{\Omega} (N_i)^\mathrm{T} (1 - \alpha_0)\frac{A}{3 \Delta t}\mathrm{tr}\mathbf{I} \zeta (2\mu \mathbf{V} + KA\mathbf{I}) \bm{B}_i^u \mathrm{d}V &\quad \text{if} \quad \mathrm{tr}[\bm{s}^\mathrm{p}] < 0 
        \end{dcases}
    \end{aligned}
\end{equation}
\begin{equation}
    \begin{aligned}
        \mathbf{K}^{up}_{ij} &= \frac{\partial \bm{r}_i^u}{\partial p_j} = 
        \begin{dcases}
        -\int_{\Omega} (\bm{B}^u_i)^\mathrm{T} \alpha \mathbf{I} N_i \mathrm{d}V  \quad &\text{if} \quad \mathrm{tr}[\bm{s}^\mathrm{p}] = 0 \\
        -\int_{\Omega}(\bm{B}^u_i)^\mathrm{T} \alpha_0 \mathbf{I} N_i \mathrm{d}V \\
        \qquad - \int_\Omega (\bm{B}^u_i)^\mathrm{T} \frac{A\mathrm{tr}\mathbf{I}(1 - \alpha_0)}{3}\zeta(2\mu \mathbf{V} + KA\mathbf{I}) N_i \mathrm{d}V \quad &\text{if} \quad \mathrm{tr}[\bm{s}^\mathrm{p}] < 0 
        \end{dcases}
    \end{aligned}
\end{equation}
\begin{equation}
    \begin{aligned}
        \mathbf{K}^{uu}_{ij} &= \frac{\partial \bm{r}_i^u}{\partial \bm{u}_j} = 
        \begin{cases}
        \int_{\Omega} (\bm{B}^u_i)^\mathrm{T} \mathbb{C}_\mathrm{dam}(d) \bm{B}^u_i \mathrm{d}V  \quad &\text{if} \quad \mathrm{tr}[\bm{s}^\mathrm{p}] = 0\\
        \int_{\Omega}(\bm{B}^u_i)^\mathrm{T} \mathbb{C}^\mathrm{t}\bm{B}^u_i \mathrm{d}V \quad &\text{if} \quad \mathrm{tr}[\bm{s}^\mathrm{p}] < 0
        \end{cases}
    \end{aligned}
\end{equation}

The Jacobian for the phase-field process is written as
\begin{equation}
\label{eq: jacobian-d}
    \begin{aligned}
        \mathbf{K}^{dd}_{ij} &= \frac{\partial r_i^d}{\partial d_j} = 
        \begin{dcases}
         \int_{\Omega} (N_i)^\mathrm{T} N_i \frac{1}{2}\frac{\partial^2 g(d)}{\partial d^2}\left[ \bm{\varepsilon}:\mathbb{C}:\bm{\varepsilon} + \frac{(1 - \alpha_0)^2}{K}p^2 + 2(1 - \alpha_0)p\mathrm{tr}\bm{\varepsilon} \right] \mathrm{d}V \\
        \qquad - \int_{\Omega} (N_i)^\mathrm{T} N_i \frac{2G_c}{\pi \ell} \mathrm{d}V + \int_\Omega \frac{2G_c \ell}{\pi}(\bm{B}_i)^\mathrm{T} \bm{B}_i \mathrm{d}V  \quad \text{if} \quad \mathrm{tr}[\bm{s}^\mathrm{p}] = 0  \\
        \int_{\Omega} (N_i)^\mathrm{T} N_i \frac{1}{2}\frac{\partial^2 g_p(d)}{\partial d^2}\epsp:\mathbb{C}:\epsp \mathrm{d}V \\
        \qquad - \int_{\Omega} (N_i)^\mathrm{T}N_i \frac{2G_c}{\pi \ell} \mathrm{d}V + \int_\Omega \frac{2G_c \ell}{\pi}(\bm{B}_i)^\mathrm{T} \bm{B}_i \mathrm{d}V  \quad \text{if} \quad \mathrm{tr}[\bm{s}^\mathrm{p}] < 0
        \end{dcases} \\
    \end{aligned}
\end{equation}

\section{Time-dependent analytical solutions for the KGD problem}
\label{sec:KGD analytical solutions}

\citet{garagash2006plane} provides the analytical solutions to the KGD problem, encompassing the evolution of pressure $p(0,t)$ and fracture aperture $w(0,t)$ at the injection point, and fracture length $L(t)$ as follows
\begin{equation}
\label{eq:p at inj point}
    \begin{aligned}
        p(0,t) =& \sigma_0 + E' \left( \frac{K'^4}{E'^4 Q t} \right)^{1/3} \left[ \frac{\pi^{1/3}}{8} + \mathcal{M} \frac{1+48\ln 2}{9\pi^{2/3}} \right] ,\\
        w(0, t) =& \left( \frac{K'^2 Q t}{E'^2} \right)^{\frac{1}{3}} \left[ \frac{1}{\pi^{1/3}} + \delta(\mathcal{M}) \frac{8(12\pi - 7 - 24\ln 2)}{9\pi^{4/3}} \right] ,\\
        L(t) =& \left( \frac{E' Q t}{K'} \right)^{\frac{2}{3}} \left[ \frac{2}{\pi^{2/3}} - \delta(\mathcal{M}) \frac{32(1+6\ln 2)}{9\pi^{5/3}} \right],
    \end{aligned}
\end{equation}
where $\sigma_0$ is the far-field minimum principal stress perpendicular to the fracture, $\delta(\mathcal{M}) = \frac{\mathcal{M}}{\sqrt{1 + 30\mathcal{M}}}$ is a small parameter proposed by~\citet{garagash2006plane}. 
The fracture length of the phase-field crack $L_\mathrm{sim}(t)$ is calculated by the following equation as~\citep{yoshioka2021variational}
\begin{equation}
    L_\mathrm{sim}(t) = \dfrac{\frac{G_c}{4c_n}\int_\mathrm{\Omega}\left[ \frac{(1 - d)^{n}}{\ell} + \ell|\nabla d|^{2} \right] \mathrm{d}V}{G_c^\mathrm{eff}}
    .
\end{equation}

\bibliographystyle{elsarticle-harv} 
\bibliography{cas-refs}

\end{document}